\documentclass[twocolumn]{aastex63}

\usepackage{natbib}
\usepackage{color}
\usepackage{graphicx}
\usepackage[caption=false]{subfig}
\usepackage{multirow}
\usepackage{float}
\usepackage{threeparttable}
\usepackage{tabularx}

\received{XXX}
\revised{YYY}
\accepted{ZZZ}
\shorttitle{Morphological, Structural and the Major Merger Evolution to $z \sim 3$}
\shortauthors{Whitney et al.}

\begin{document}

\title{Galaxy Evolution in all Five CANDELS Fields and IllustrisTNG: Morphological, Structural, and the Major Merger Evolution to $z \sim 3$}

\author{A. Whitney}
\affil{University of Nottingham, School of Physics \& Astronomy, Nottingham, NG7 2RD, UK}
\author{L. Ferreira}
\affiliation{University of Nottingham, School of Physics \& Astronomy, Nottingham, NG7 2RD, UK}
\author{C. J. Conselice}
\affiliation{University of Nottingham, School of Physics \& Astronomy, Nottingham, NG7 2RD, UK}
\affiliation{Jodrell Bank Centre for Astrophysics, University of Manchester, Oxford Road, Manchester UK}
\author{K. Duncan}
\affiliation{SUPA, Institute for Astronomy, Royal Observatory, Blackford Hill, Edinburgh, EH9 3HJ, UK}
\affiliation{Leiden Observatory, Leiden University, PO Box 9513, NL-2300 RA Leiden, The Netherlands}

\correspondingauthor{Amy Whitney}
\email{amy.whitney@nottingham.ac.uk}



\begin{abstract}

A fundamental feature of galaxies is their structure, yet we are just now understanding the evolution of structural properties in quantitative ways. As such, we explore the quantitative non-parametric structural evolution of 16,778 galaxies up to $z\sim3$ in all five CANDELS fields, the largest collection of high resolution images of distant galaxies to date. Our goal is to investigate how the structure, as opposed to size, surface brightness, or mass, changes with time. In particular, we investigate how the concentration and asymmetry of light evolve in the rest-frame optical. To interpret our galaxy structure measurements, we also run and analyse 300 simulation realisations from IllustrisTNG to determine the timescale of mergers for the CAS system. We measure that from $z=0-3$, the median asymmetry merger timescale is $0.56^{+0.23}_{-0.18}$Gyr, and find it does not vary with redshift. Using this data, we find that galaxies become progressively asymmetric at a given mass at higher redshifts and we derive merger rates which scale as $\sim(1+z)^{1.87\pm0.04}$Gyr$^{-1}$, which agrees well with recent machine learning and galaxy pair approaches, removing previous inconsistencies. We also show that far-infrared selected galaxies that are invisible to \textit{HST} have a negligible effect on our measurements. We also find that galaxies are more concentrated at higher redshifts. We interpret this as a sign of how their formation occurs from a smaller initial galaxy that later grows into a larger one through mergers, consistent with the size growth of galaxies from `inside-out', suggesting that the centres are the oldest parts of most galaxies.

\vspace{10mm}
\end{abstract}



\section{Introduction} \label{sec:intro}

Throughout the history of galaxy studies, the most common way to derive galaxy evolution is through examining some property as a function of time. This famously includes the evolution of star formation, stellar mass, metallicity, and other properties. One of the most fundamental properties that we are still exploring is the morphological or structural evolution of galaxies, which is an integrated result of the many different galaxy properties and formation processes \citep[e.g.,][]{conselice08, mortlock13, conselice14, huertas16}.

There are many ways in which to trace the structural evolution of galaxies. The most simplistic and direct way is investigating the size evolution \citep[e.g.,][]{trujillo07, buitrago08, allen17, whitney19}, and the surface brightness evolution \citep[e.g.,][]{whitney20}, as well as simply the evolution of apparent morphology classified into Hubble types/peculiars \citep[e.g.,][]{conselice05}. Another way to examine the evolution of galaxy structure is to examine the bulge and disk components of galaxies and how these evolve together \citep{bruce14, margalef18}. What has not been carried out in any detail is the quantitative evolution of galaxies as measured with non-parametric parameters. These parameters, including concentration and asymmetry \citep{kent85, conselice03}, reveal the processes of galaxy assembly through the systematic change of galaxy light over time.

Throughout a galaxy's lifetime, it will undergo several processes that will alter its structure and its morphology.  Within a cosmological context of $\Lambda$-CDM, this includes the formation of bulges and then disks.  In the simplest paradigm galaxies collapse into small systems that grow through star formation and mergers with other galaxies.  At some point gas accretion will also occur and this is a primary method by which spiral arms and disks are formed.  This process includes the accumulation of gas forming into stars that will expand galaxies in their outer parts \citep{whitney19}. This also includes mergers that will lead to structural peculiarities, and eventually, for some, into more concentrated systems. Furthermore, within clusters of galaxies processes such as ram pressure stripping \citep{gunn72}, starvation \citep{larson80}, and harassment \citep{moore96} are all tied to galaxy star formation history and can strongly influence the physical and morphological properties of a galaxy. 

A galaxy's morphology is traditionally defined as the point at which it lies on the `tuning fork' diagram, first described by Edwin Hubble \citep{hubble26}, whereby galaxies are defined as either spirals or ellipticals/S0s. Historically, these morphological classifications were done visually \citep[e.g.,][]{devaucouleurs59, sandage75, vandenbergh76, lintott08, lintott11}, whereby an individual examines galaxy images and assigns labels to those images based in their visual appearance. However, this method gives rise to errors and biases and the sheer sample size of current and upcoming surveys mean this is quickly becoming inefficient even for large citizen-science projects such as Galaxy Zoo \citep{cheng20}. In order to remove some of these problems it is important to use a less subjective and more quantitative way of classifying galaxies. One such method is a non-parametric system that seeks to measure the concentration, asymmetry, clumpiness, Gini and $M_{20}$ (CAS parameters) of galaxies by using measured light distribution. This system is described in papers such as \cite{conselice00a, conselice02, lotz04}. Using this method, galaxies can be placed in parameter space and from this, it can be seen that all major classes of galaxies in various phases of evolution are easily differentiated \citep{conselice03}. Furthermore, classical classifications of galaxies are unable to be used at higher redshifts, whereby most galaxies are not elliptical or spirals \citep[e.g.,][]{conselice05, mortlock13}.

An important stage of galaxy evolution that can be measured with these parameters is when a galaxy undergoes a merger. Mergers can be identified using methods such as pair fractions \citep{man16, mundy17, duncan19, ventou19}, deep learning models \citep{ferreira20}, and by using the CAS parameter space \citep{conselice03, lotz04, conselice08, lopezsanjuan09}. Major mergers lie in a specific areas of these parameter spaces and as such, they are a useful tool in determining whether a galaxy is a merger or not. The merger rate can then be calculated from this to determine the role of mergers in forming galaxies.  

In this paper we investigate the general evolution of galaxy structure through cosmic time. We start with visual estimates of morphology and structure and then we examine the quantitative structural evolution of these systems. We use IllustrisTNG simulations to help us interpret these structures from which we derive the rates of galaxy formation processes such as those driven by mergers and how light concentration in a galaxy changes with time.

Throughout this paper we use AB magnitudes and assume a $\Lambda$-CDM cosmology with H$_0$ = 70 kms$^{-1}$Mpc$^{-1}$, $\Omega_m$ = 0.3, and $\Omega_{\Lambda}$ = 0.7.

\section{Data \& Methods}

\subsection{Data}

\begin{figure}[!ht]
\includegraphics[width=0.475\textwidth]{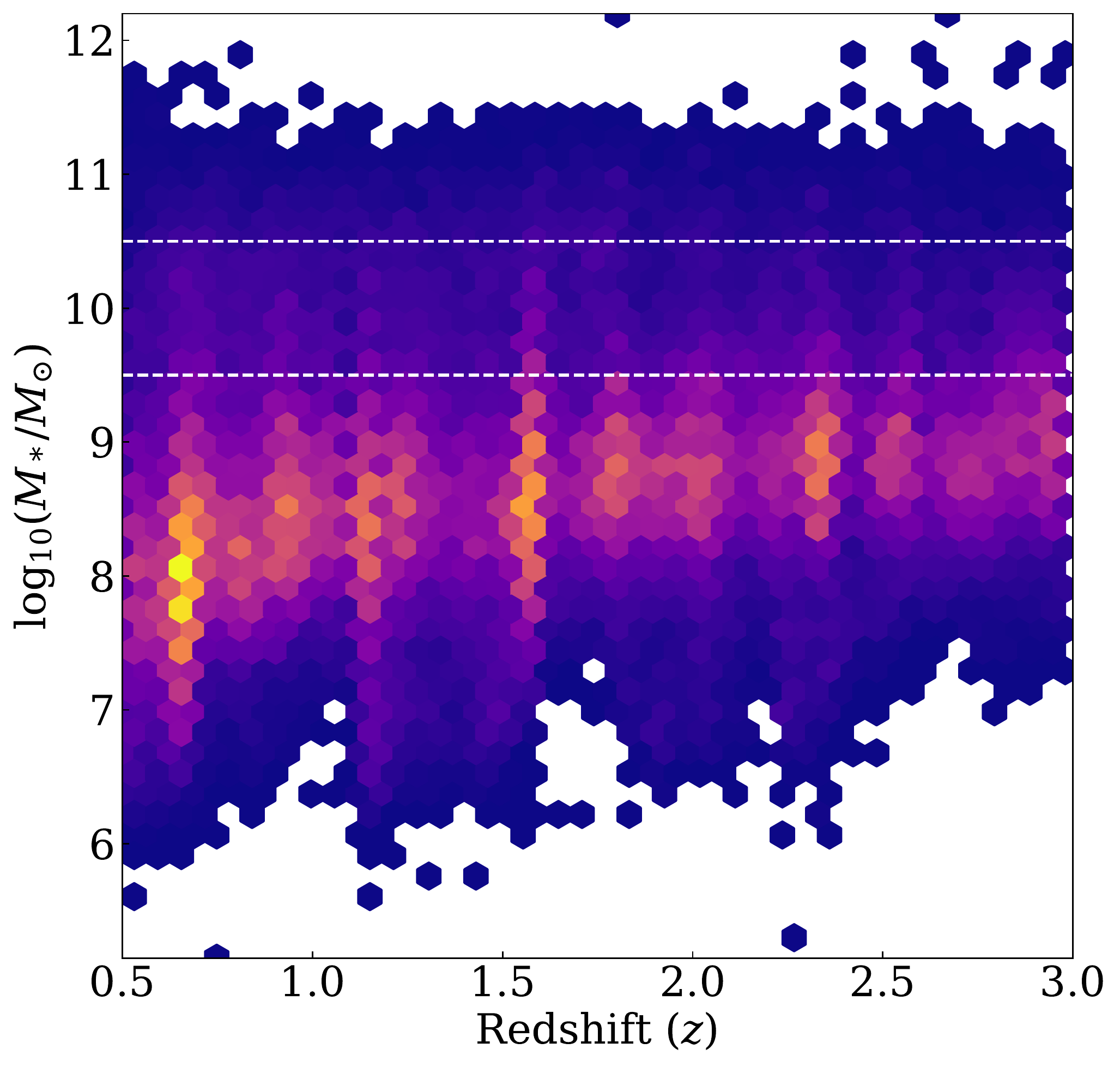}
\caption{Mass and redshift distribution of the sample of galaxies. Yellow indicates a higher density of galaxies and purple indicates the lowest density. White areas indicates there are no galaxies are present in this region of parameter space. The dashed horizontal lines at log$_{10}(M_{*}/M_{\odot})$ = 9.5 and log$_{10}(M_{*}/M_{\odot})$ = 10.5 indicate the boundaries between the two mass bins used in this study.}
\label{fig:distribution}
\end{figure}

We use a sample of 16,778 galaxies at redshifts in the range 0.5 $< z <$ 3 with stellar masses from $10^{9.5}M_{\odot} < M_{*} <  10^{12.2}M_{\odot}$ and only select galaxies with S/N $>$ 10. This signal-to-noise cut removes only 340 galaxies from the initial sample. The mass and redshift distribution can be seen in Figure \ref{fig:distribution} where the yellow regions indicate a higher density of galaxies and purple indicates a lower density of galaxies. We select two samples from the 16,778 galaxies; a low mass sample with $10^{9.5}M_{\odot} < M_{*} <  10^{10.5}M_{\odot}$ and a high mass sample with $M_{*} >  10^{10.5}M_{\odot}$.  We later also consider how our results change if we used a constant co-moving number density selected sample.  The mass limits for our mass-selected sample are shown as horizontal dashed lines on Figure \ref{fig:distribution}. Masses and redshifts are determined using the method described in \cite{duncan19}. A brief description of the process used to obtain the measurements is given in \S\ref{sec:z_M}.

This study makes use of data from the Cosmic Assembly Near-infrared Deep Extragalactic Survey \citep[CANDELS;][]{grogin11, koekemoer11}. This survey covers 800 arcmin$^2$ over five fields; GOODS-North, GOODS-South, COSMOS, UDS, and EGS. Our sample consists of galaxies from each of the five fields. CANDELS makes use of the Advanced Camera for Surveys (ACS) and the Wide Field Camera 3 (WFC3) on the \textit{Hubble Space Telescope} (\textit{HST}). For this work we use imaging data from the F814W, F125W, and F160W filters. This filters will be referred to as $I_{814}$, $J_{125}$, and $H_{160}$ from this point onward. 

We measure all parameters in the optical rest-frame at approximately $\lambda_{rest} \sim 4000${\AA}.  We do this by locating the observed filter where  $\lambda_{rest} \sim 4000${\AA} falls as determined by its redshift.  We then use that filter in our further analysis to have a consistent rest-frame optical view of our galaxies.  The idea here is to measure everything at a constant optical rest-frame wavelength, to avoid structural biases produced by changes in rest-frame wavelength.   The rest-frame wavelength and associated filter are given for each redshift bin in Table \ref{tab:morph_filters}.

\begin{table}[!ht]
\centering
\caption{The bands corresponding to the rest-frame optical at each redshift.}
\begin{threeparttable}
  \begin{tabularx}{0.475\textwidth}{*3{>{\centering\arraybackslash}X}}
  \hline 
  \hline
  $z$ & $O_{i,j}^{raw}$ & $\lambda_{rest}$\\
  \hline
  0.75 & $I_{814}$ & 4650{\AA} \\
  1.25 & $I_{814}$ & 3620{\AA} \\
  1.75 & $J_{125}$ & 4550{\AA} \\
  2.25 & $J_{125}$ & 3850{\AA} \\
  2.75 & $H_{160}$ & 4270{\AA} \\
  \label{tab:morph_filters}
\end{tabularx}

\begin{tablenotes}
  \small
  \item \textbf{Note.} Column 1 gives the midpoint of the redshift bin. Each redshift bin spans a redshift range of $\Delta z = 0.5$. Column 2 gives the band corresponding to the optical rest-frame and column 3 gives the rest frame wavelength probed.
\end{tablenotes}
\end{threeparttable}

\end{table}

We use visual morphological classifications from \cite{kartaltepe15} who base the majority of their classifications on $H$-band images, however $J$- and $V$-band images are also used for some features such as clumpy light.   For a summary of how these classifications are done and in which morphological classes they are placed, see \cite{kartaltepe15}.

\subsection{\textsc{Morfometryka}}\label{sec:morf}

We measure the non-parametric concentration, asymmetry, and clumpiness parameters using the \textsc{Morfometryka} code \citep{ferrari15}. These measurements are made within the Petrosian region \citep[e.g.,][]{conselice00a}. This is defined as the area with the same axis ratio and position angle as the galaxy and with major axis equal to $N_{R_{\textup{Petr}}}R_{\textup{Petr}}$, where $N_{R_{\textup{Petr}}} = 1.5$ and $R_{\textup{Petr}}$ is the Petrosian radius. Below we give a description of each index we use, although for more details see \cite{conselice03}.

\subsubsection{Asymmetry}

The asymmetry, $A$, is calculated in the same way as in \cite{conselice03} whereby a galaxy image rotated by 180$^{\circ}$ is subtracted from the original source image, and the absolute value of the total light in this self-subtracted image is divided by the total light in the original image. Asymmetry through various tests has been shown to be one of the most robust non-parametric structural parameters to measure and use in analyses. The formula for calculating the asymmetry index ($A$) we use is outlined in detail in \cite{conselice03} and is given by:

\begin{equation}
    A = {\rm min} \left(\frac{\Sigma|I_{0}-I_{180}|}{\Sigma|I_{0}|}\right) - A_{bkg}
\end{equation}

\noindent where $I_{0}$ represents the original galaxy image, and  $I_{180}$ is this image after rotating it by 180 deg from its centre. The asymmetry value is calculated through an iterative approach to find the centre of the rotation which is altered to find the one that gives the minimum asymmetry value.  This minimum asymmetry which is used as the final asymmetry value \citep[e.g.,][]{conselice00a} with a search radius typically a pixel or half pixel. $A_{bkg}$ is the background asymmetry of the image. This background term differs from the original application of the CAS parameter measurements whereby the background asymmetry is determined from a single region. \textsc{Morfometryka} initially did not include such a background correction. In this case we use the same method for determining $A_{bkg}$ as in \cite{tohill20} whereby a $10 \times 10$ grid is overlayed on the image in an area outside of the galaxy segmentation map, the asymmetry of the individual cells are measured and then the median of these values is taken for $A_{bkg}$. This removes the bias present in the original method of measuring the asymmetry of a single background region and ensures the measurement is more robust. 

\subsubsection{Concentration} \label{sec:conc}

The concentration of a galaxy is used to distinguish between the different types of galaxies; for example, early type galaxies and their immediate progenitors tend to have a higher concentration than late type galaxies \citep[e.g.,][]{bershady00, conselice03}. Concentration, $C$, is defined as the ratio between two circular radii containing certain fractions of the total flux of the galaxy \citep{kent85}. We use $R_{20}$, the radius containing 20\% of the total light, and $R_{80}$, the radius containing 80\% of the total light. Therefore, the concentration is given as 

\begin{equation}
    C = 5 \times \textup{log}_{10}\left(\frac{R_{80}}{R_{20}}\right).
    \label{eq:C}
\end{equation}

\noindent This calculation of the concentration is sensitive to seeing effects that are more prominent in the central regions and is therefore sensitive to the value of the inner radii $R_{20}$ \citep{ferrari15}.  We however, investigate other forms of concentration indices, and fully explore the idea of light concentration and what it actually implies for our galaxies.  We also consider the $R_{50}$ and $R_{90}$ radii to avoid problems with the inner parts of galaxies, as well as examine the Petrosian radii measures of light concentration.

\subsubsection{Clumpiness} \label{sec:clumpiness}

The clumpiness, $S$, is a measure of the small scale structure within a galaxy. A higher clumpiness indicates that there are clumps of material within a galaxy, for example spiral galaxies contain many star forming regions and therefore contain many clumps of material. Elliptical galaxies on the other hand are generally smooth and therefore have a clumpiness that is close to zero. $S$ is calculated in the same way as in \cite{conselice03} whereby the image is first smoothed by a filter and then subtracted from the original image. The flux contained within this residual is then divided by the flux contained with the original image. In this case, the filter used to smooth the original image is a Hamming window of size $R_{\textup{Petr}}$/4. Formally, the clumpiness can be described by the following

\begin{equation}
    S = \sum_{i,j}\frac{|I(i, j)-I_{S}(i, j)|}{|I(i, j)|}-S_{bkg}
\end{equation}

\noindent where $I(i, j)$ is the original image, $I_{S}(i, j)$ is the smoothed image, and $S_{bkg}$ is the average smoothness of the background. 

\subsection{Correcting CAS Values for Redshift Effects} \label{sec:corrections}

Galaxy structure will change for a given galaxy when that same galaxy is viewed at higher redshifts.  To decouple evolution from this effect, we need to account for this.  In order to correct for redshift effects, we artificially redshift a sample of galaxies at 0.5 $< z <$ 1 in redshift intervals of 0.5 to a maximum redshift of $z = 2.75$. We take the galaxies in this initial bin from both GOODS fields and consider this our fiducial sample. We simulate these galaxies to higher redshifts and in the other CANDELS fields. We do this using the method described in \cite{tohill20}. This method considers a number of effects. Firstly, it considers the rebinning factor $b$, which is the decrease in the apparent size of the galaxy when viewed at a higher redshift. This is done by following the method outlined in \cite{conselice03} and \cite{ferreira18}. Luminosity evolution is also considered and this is implemented in the form found in \cite{whitney20} whereby the intrinsic surface brightness goes as 

\begin{equation}
    \mu_{\textup{int}} \propto (1+z)^{-0.18}
\end{equation}

\noindent for a mass-selected sample of galaxies. After applying the rebinning factor, cosmological dimming of the form $(1 + z)^{-4}$, as found by \cite{tolman30}, is applied. Finally, the image is convolved with the PSF corresponding to the rest-frame filter and inserted into an actual CANDELS background. For the first redshifting interval ($z = 0.75$ to $z = 1.25$) we do not convolve with the PSF as the filter corresponding to the optical rest-frame is the same for both redshifts. We then use \textsc{Morfometryka} to measure the CAS parameters at each redshift and compare these new values to the parameters measured at the original redshift. The differences are then applied to the real galaxies at the corresponding redshift. This change is dependent on the field due to the depth reached by each of the fields. The GOODS fields reach a greater depth than the COSMOS, EGS, and UDS fields and as such, we apply a different correction to each field. We also ensure that we are comparing the same galaxies at each redshift interval.  To do this we only consider at all redshifts those galaxies that can be measured in the highest redshift as this ensures that we are comparing exactly the same galaxies at all redshifts to create a fair comparison and relevant correction factors.

\begin{figure*}[!ht]
\centering
\begin{tabular}{cc}
\subfloat{\includegraphics[width = .49\textwidth]{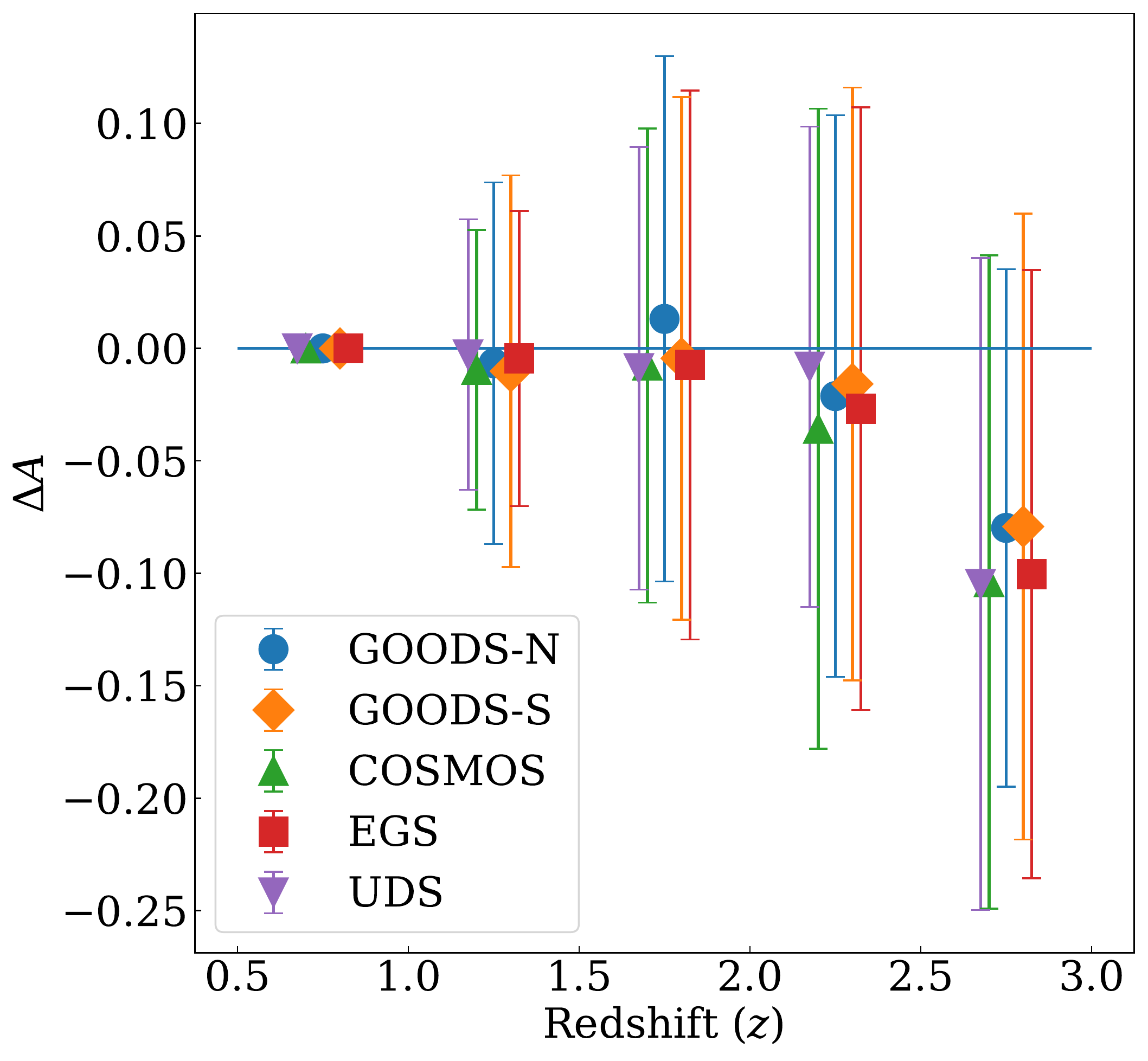}} &
\subfloat{\includegraphics[width = .49\textwidth]{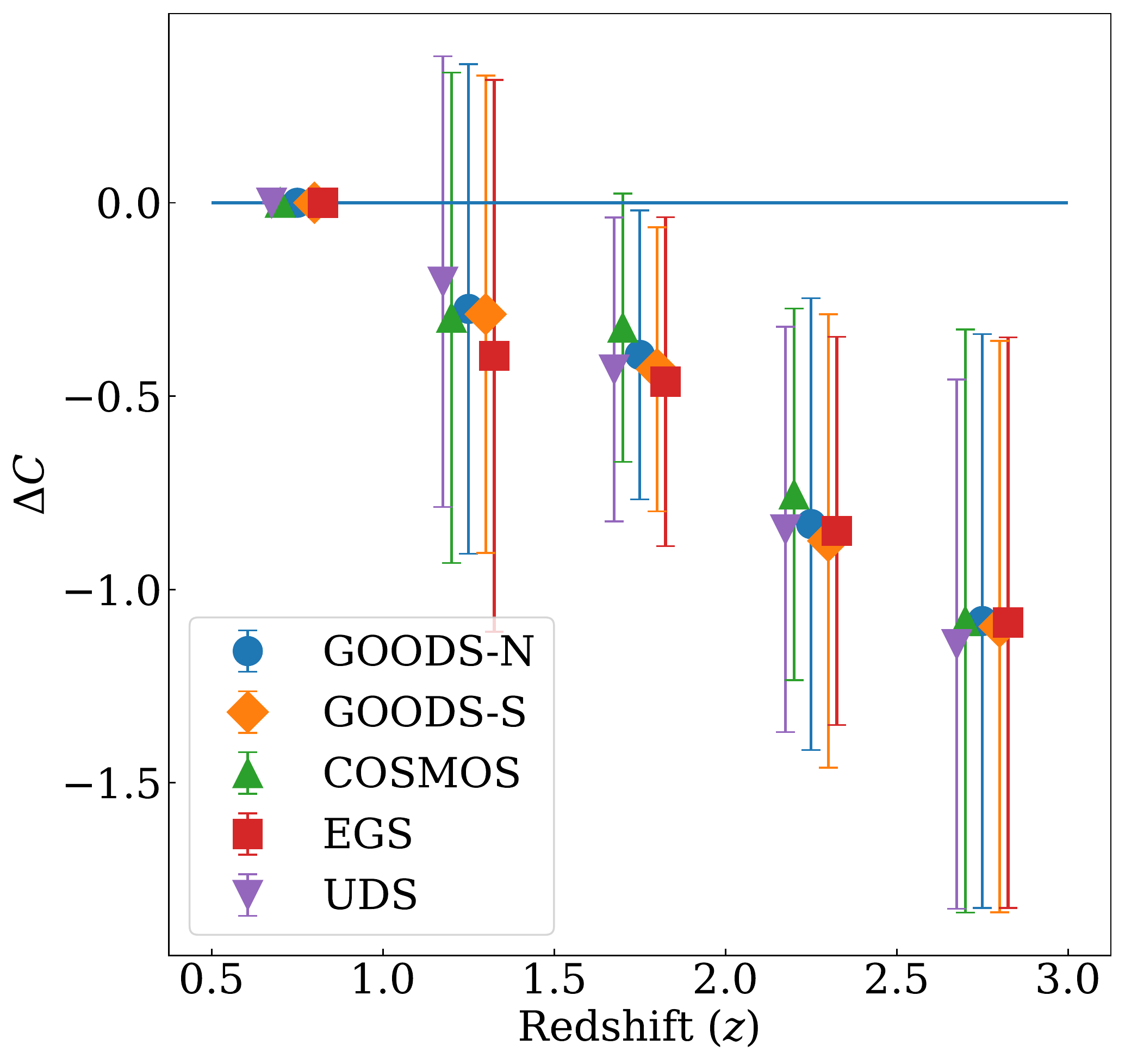}} \\
\end{tabular}
\caption{Left: mean asymmetry correction applied to all five CANDELS fields. Right: mean concentration correction for all CANDELS fields. The error bars are equal to one standard deviation from the mean. In both panels, the GOODS-North correction is shown by a blue circle, GOODS-South by an orange diamond, COSMOS by a green triangle, EGS by a red square, and UDS by a purple inverted triangle.}
\label{fig:z_corr}
\end{figure*}

The left panel of Figure \ref{fig:z_corr} shows the average asymmetry corrections applied to all CANDELS fields. The right panel of Figure \ref{fig:z_corr} shows the average concentration corrections applied to the five CANDELS fields. Table \ref{tab:corrections} gives the values of these corrections for each redshift bin within each field. 

\begin{table*}
\centering
\caption{The asymmetry and concentration corrections applied to each field at each redshift interval.}
  \begin{tabular}{ccccccccccc}
  \hline 
  \hline
    & \multicolumn{5}{c}{Asymmetry} & \multicolumn{5}{c}{Concentration} \\
  $z$ & GOODS-N & GOODS-S & COSMOS & EGS & UDS & GOODS-N & GOODS-S & COSMOS & EGS & UDS\\
  \hline
  0.75 & 0 & 0 & 0 & 0 & 0 & 0 & 0 & 0 & 0 & 0 \\
  1.25 & -0.010 & -0.010 & -0.009 & -0.004 & -0.007 & -0.27 & -0.29 & -0.30 & -0.40 & -0.20 \\
  1.75 & 0.008 & -0.004 & -0.008 & -0.007 & -0.023 & -0.39 & -0.43 & -0.32 & -0.46 & -0.43 \\
  2.25 & -0.032 & -0.016 & -0.036 & -0.027 & -0.060 & -0.83 & -0.87 & -0.75 & -0.85 & -0.84 \\
  2.75 & -0.084 & -0.079 & -0.104 & -0.100 & -0.118 & -1.08 & -1.10 & -1.08 & -1.09 & -1.14 \\
  \label{tab:corrections}
  \end{tabular}

\end{table*}

\subsection{Photometric Redshifts and Stellar Masses} \label{sec:z_M}

The photometric redshifts of the galaxies within our sample are calculated using the method described in \cite{duncan19}. The photometric redshift software \textsc{eazy} \citep{brammer08} is used to determine the template-fitting estimates and three separate template sets are used and fit to all available photometric bands. The templates used include zero-point offsets to the input fluxes and additional wavelength-dependent errors. A Gaussian process code (GPz; \cite{almosallam16}) is then used to calculate further empirical estimates using a subset of the available photometric bands. Individual redshift posteriors are calibrated and the four estimates are combined in a statistical framework via a hierachical Bayesian combination to produce a final redshift estimate. For a more in-depth description of the process, see section 2.4 of \cite{duncan19}.

The galaxy stellar masses we use are measured by using a modified version of the spectral energy distribution (SED) code described in \cite{duncan14}. Instead of finding the best-fit mass for a fixed input redshift, the stellar mass is estimated at all redshifts in the photo-$z$ fitting range. Also included in these estimates is a so-called `template error function' (the method for this is described in \cite{brammer08}) to account for uncertainties introduced by the limited template set and any wavelength effects. 

This mass-fitting technique uses \cite{bruzual03} templates and includes a wide range of stellar population parameters and assumes a \cite{chabrier03} initial mass function. The assumed star formation histories follow exponential $\tau$-models for both positive and negative values of $\tau$. Characteristic timescales of $\left|\tau\right|$ = 0.25, 0.5, 1, 2.5, 5, and 10 are used, along with a short burst ($\tau$ = 0.05) and continuous star formation models ($\tau \gg$ 1/$H_0$).

We compare the mass measurements we make to the average of those determined by the several teams within the CANDELS collaboration \citep{santini15}. This is done in order to ensure that the stellar mass estimates do not suffer from systematic biases. There is some scatter between the two mass estimates, however our mass estimates are not affected by any significant biases compared to others. For further details on the method and models used, see \S 2.5 of \cite{duncan19} for an extensive discussion of the masses we use here.

\subsection{IllustrisTNG Simulations}\label{subsec:tng}

To interpret our results, we  investigate how non-parametric morphologies change with redshift in a controlled manner with simulated galaxies. We do this as we want to disentangle real evolution effects from redshift effects. Additionally, by measuring these structural properties in simulated data we can estimate the observability timescales of galaxy mergers since it is possible to follow the time evolution of particular galaxies in this way \citep[e.g.,][]{lotz08}. With this goal in mind, we use data from the IllustrisTNG simulations \citep{Pillepich2018b, Springel2018, Nelson2018a, Naiman2018, Marinacci2018, Nelson2019, Nelson2019b, Pillepich2019}, which is a suite of cosmological, gravo-magnetohydrodynamical simulation runs, ranging within a diverse set of particle resolutions for three comoving simulation boxes of size, $50, 100, 300 \rm \ Mpc \ h^{-1}$, named TNG50, TNG100 and TNG300, respectively.

Here we use data from both TNG50-1 and TNG300-1. For estimating the merging timescales we focus in the largest simulation box TNG300-1, as it provides us with a more mass complete sample at higher redshifts. However, when doing direct comparisons between the morphology measured in CANDELS and in the simulations we use data from TNG50-1 as its mass and spatial resolution produce more realistic morphologies.  This enable us to generate images of simulated galaxies that are embedded in a cosmological context.

To measure the observability timescales of pair galaxy mergers discussed in Section \S \ref{subsec:timescales_tng}, we select galaxies using the TNG300-1 merger trees by searching and locating galaxies in the simulation that have had only one major merger event within redshifts $z = 2$ to $z \sim 0$.   This is to avoid contamination in the structure from past merger events.  To balance any potential issues with the mass resolution of TNG300-1, we limit our analysis only to massive galaxies with $M_{*} > 10^{10} M_\odot$, which are in general represented by thousands of stellar particles. For these galaxies, we select all the snapshots and subhalos that are $\pm \ 2 \rm \ Gyr$ of the snapshot from where the merger event takes place. We then narrow down our selection to 300 distinct galaxies, each with 35 different snapshots, resulting in a total of $\sim 10,000$ distinct objects. This ensures that we extensively probe not only around the merging event, but also the stages where the galaxies does not show signs of merging. 

For a comparison between the non-parametric morphology and merger classifications of CANDELS galaxies and simulated galaxies we use the small box, high mass resolution TNG50-1 simulation, which is capable of producing output images in higher resolutions due to the high mass resolution and smaller gravitational softening length. Here we do two different selections on TNG50-1 galaxies, which are described in \S \ref{subsec:compare_to_sim}.

The IllustrisTNG data contains the information from the stellar, gas, and dark matter particles for each source. However, to create mock broadband images from this information we post-process each stellar particle with a population synthesis process, as each particle represents a large region that can be described by a rich stellar population based on its age, mass, and metallicity. Instead of using the approach outlined in \cite{ferreira20}, we follow the recipes from \cite{Trayford2017OpticalSKIRT} and \cite{  Vogelsberger2020High-redshiftFunctions} to post-process the simulation data with the Monte Carlo dusty radiative transfer code SKIRT \citep{SKIRT8, Camps2020SKIRTGrains}. We also include the resampling of the star-forming regions outlined in \cite{CAMPS2016} and \cite{Trayford2017OpticalSKIRT}, as this is particularly important to avoid problems with the coarse representation of star forming regions. For each source simulated with SKIRT we produce observations in four different orientations in the 3D volume containing the galaxy cutout. Three of those are aligned with the simulation box axis: xy, xz and yz. We also include a fourth orientation covering an octant of the 3D volume. This post-processing step is independent of the simulation run used, the only difference being the number of stellar and gas particles available and the output size of the datacube.

The output from SKIRT is a datacube with a spectral energy distribution (SED) for each pixel, which is then convolved with the filter response function for the \textit{HST} CANDELS filters used in this work, namely $I_{814}$, $J_{125}$, and $H_{160}$. The resulting broadband images are stored in two ways: without any observation effects and with \textit{HST}-matched properties, including PSF, noise level, and sky background.  The final images are then processed with \textsc{Morfometryka} to measure their CAS values. We later use these simulation outputs as a method for understanding our observational results.  

\section{Results}

In the following subsections we discuss the results of our study in terms of the morphological and structural evolution of galaxies to $z = 3$.  We first describe the visual morphological classification evolution of our sample of galaxies, while later we discuss the quantitative evolution using the CAS parameters. Finally we use these results to derive the merger fraction and merger rate evolution for our sample and the resulting number of mergers and mass accretion from mergers over this cosmic time.

\subsection{Visual Classifications}

\begin{figure}[!ht]
\centering
\includegraphics[width=0.475\textwidth]{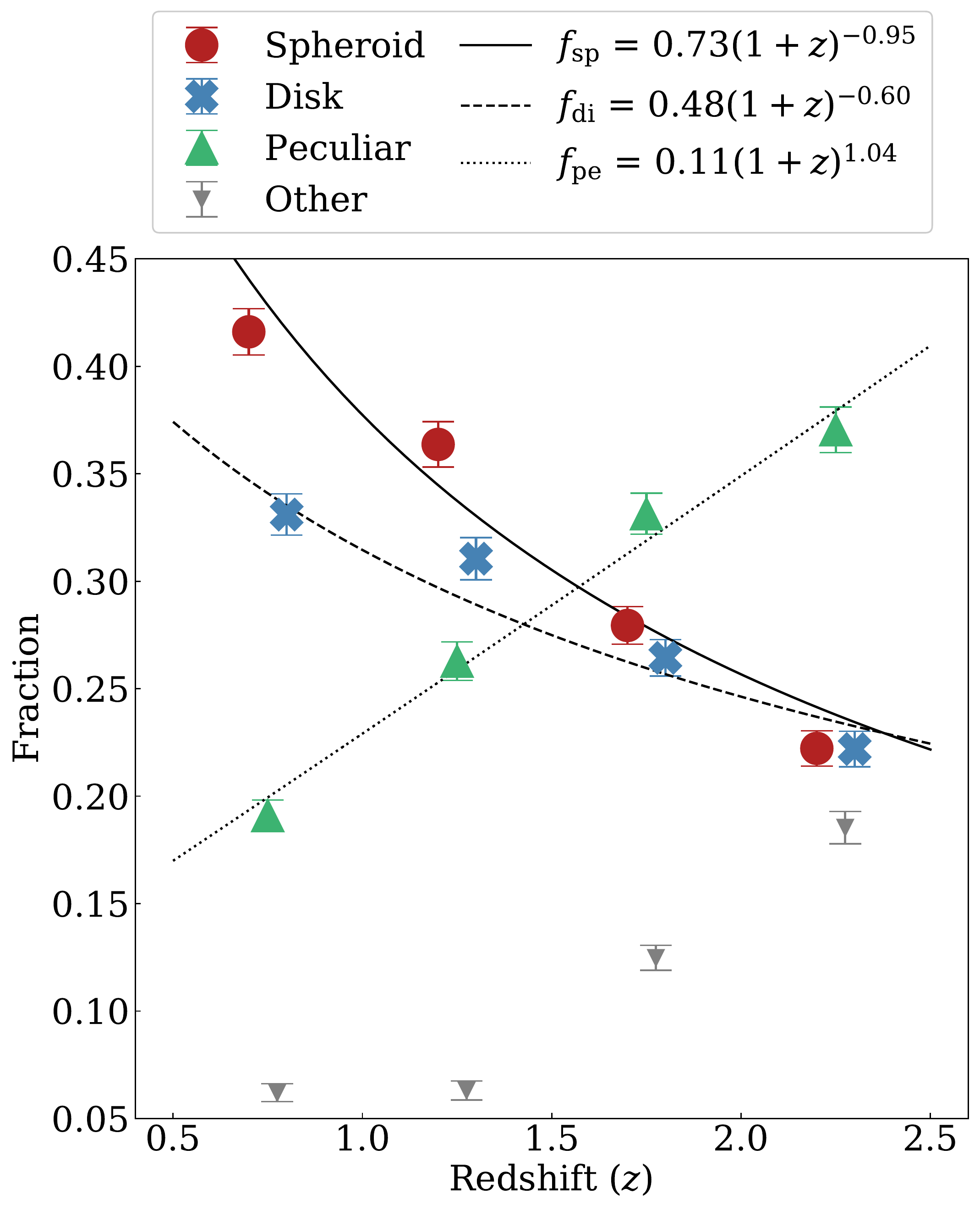}
\caption{Evolution of the fractions of galaxies in our sample within a given visually morphological classification. Spheroid galaxies are shown as red red circles, disk galaxies as blue crosses, peculiar galaxies as green diamonds, and unclassifiable objects as grey inverted triangles. Also shown are the power law fits to the spheroid (solid line), disk (dashed line), and peculiar categories (dotted line).  For the error bars on these fitted parameters see the text.}
\label{fig:fractions}
\end{figure}

Using the visual classifications from the CANDELS survey from \cite{kartaltepe15}, we examine the fraction of each galaxy type within our sample. We consider those galaxies that are classified as the following: spheroid, disk, peculiar, and other. A galaxy is considered to have a classification if the fraction of classifiers within \cite{kartaltepe15} that deem it to be that particular type of object is greater than 0.6, and the fraction of all other classifications is less than 0.6. For peculiar galaxies, we consider any galaxy where the fraction of classifiers that consider that galaxy to be irregular in shape is greater than 0.6; no other conditions are required for this group. The other group consists of galaxies for which no consensus could be reached as to its morphological type, galaxies that were deemed unclassifiable, or galaxies that had no classification given. Figure \ref{fig:fractions} shows the evolution of the fraction of galaxies in each of these groups within our sample across the redshift range $0.5 < z < 2.5$. The spheroid galaxies are shown as red circles, disk galaxies as blue crosses, and peculiar galaxies as green triangles. The unclassifiable objects are shown as grey inverted triangles. We also fit power laws of the form $\alpha(1+z)^{\beta}$ to the spheroid (solid line), disk (dashed line), and peculiar (dotted line) categories. We find that the fractions evolve as:

\begin{equation}
    f_{\textup{sp}} = (0.73 \pm 0.09) (1+z)^{-0.95 \pm 0.16}
\end{equation}

\noindent for the spheroid galaxies, 

\begin{equation}
    f_{\textup{di}} = (0.48 \pm 0.05) (1+z)^{-0.60 \pm 0.13}
\end{equation}

\noindent for the disk galaxies, and 

\begin{equation}
    f_{\textup{pe}} = (0.11 \pm 0.01) (1+z)^{1.04 \pm 0.10}
\end{equation}

\noindent for the peculiar galaxies. The fraction of both disk and spheroid galaxies increase with cosmic time, whereas the fraction of peculiar galaxies decreases as redshift decreases. Note that these fits are only valid down to $z \sim 0.5$.  It is clear that extrapolating these to $z \sim 0$ would over-predict the number for each type. The fraction of `other' galaxies evolves from $\sim$20\% at $z \sim 2.25$ and decreases to $\sim$6\% at $z \sim 0.75$ due to the fact that galaxy features will be more distinguishable at lower redshifts due to increased resolution and fewer effects such as surface brightness dimming. 

\subsection{Asymmetry and Concentration}

We next explore the data by comparing the measured asymmetry and concentration values with the visual classifications from \cite{kartaltepe15} as described in the previous section.  In this section we examine those galaxies that are identified as disk, spheroid, or peculiar galaxies.  We first examine the uncorrected CAS parameters here, and later investigate a way to correct these parameters for redshift effects.  Essentially, the uncorrected values are comparable to each other at a given redshift (internally consistent), whereas the corrected values are comparable across all redshifts (externally and internally consistent).  

\begin{figure*}[!ht]
\includegraphics[width=\textwidth]{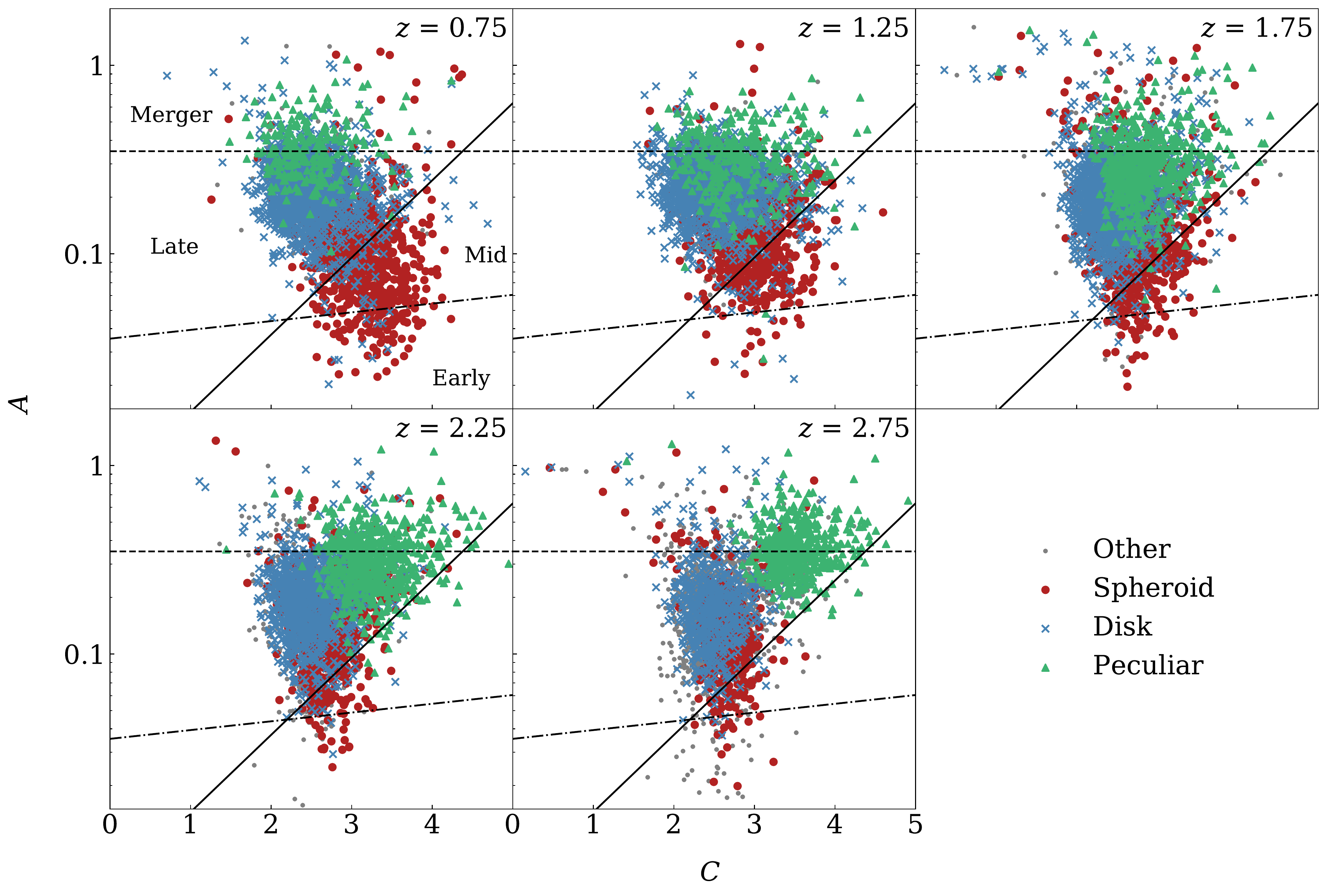}
\caption{Evolution of the concentration-asymmetry plane with redshift, colour-coded by visual classification. Red circles are galaxies classified as spheroids, blue crosses as disks, and green triangles as peculiar galaxies. Any galaxies that do not fit within these three classifications are shown as small grey circles. The lines are boundaries between galaxy types derived from \cite{bershady00}. Galaxies that are above the dashed line are considered to be mergers. The dot-dashed line is the boundary between early- and intermediate-type galaxies and the solid line is the boundary between late- and intermediate-type galaxies. The redshift indicated in the top right corner of each panel is the midpoint of the redshift bin whereby the upper and lower limits are 0.25 either side of this point.}
\label{fig:vccva}
\end{figure*}

In Figure \ref{fig:vccva}, we show the evolution of the uncorrected concentration versus asymmetry for the galaxies within the sample that have one of our three main classifications. Red circles indicate a spheroid galaxy, blue crosses a disk galaxy, and green triangles a peculiar galaxy. Galaxies that fit none of these categories are shown as small grey circles. Also shown are the cuts between galaxy types taken from \cite{bershady00}; the solid black line is the boundary between intermediate- and late-type galaxies whereby the galaxies that lay above this line are considered to be late-type galaxies, the dot dashed line is the boundary between early- and intermediate-type galaxies whereby galaxies below this line are considered to be early-type galaxies, and the dashed line is the boundary between mergers and non-mergers whereby galaxies that lie above this line are mergers. 

At all redshifts, galaxies classified as spheroids tend to have a higher concentration and lower asymmetry, than the two other classifications. Whereas peculiar galaxies have a higher asymmetry but lower concentration. This is consistent with the results found by \cite{conselice03} for nearby galaxies. The range of concentration values increases with time but the overall distribution of these values for the spheroid and disk galaxies remains roughly constant with redshift. The asymmetry of the peculiar galaxies is centred on the $A >$ 0.35 merger condition, however, the mean concentration changes from $\langle C \rangle$ = 3.52 at $2.5 < z < 3.0$ to $\langle C \rangle$ = 2.49 at $0.5 < z < 1.0$. The mean values for spheroid galaxies are $\langle C \rangle$ = 2.67 and $\langle A \rangle$ = 0.16 at the highest redshift. These values change to $\langle C \rangle$ = 3.10 and $\langle A \rangle$ = 0.12 at the lowest redshift. Galaxies classified as being disks have average concentration and asymmetry values of $\langle C \rangle$ = 2.52 and $\langle A \rangle$ = 0.20 at $2.5 < z < 3.0$ and these values do not change significantly within the lowest redshift bin of $0.5 < z < 1.0$ where $\langle C \rangle$ = 2.50 and $\langle A \rangle$ = 0.23. Thus the disks change their value the least and the spheroids the most. These results also show that galaxy classification is consistent and can be carried out to $z \sim 3$.

\subsubsection{Disk/Bulge Dominated}

\begin{figure*}[!ht]
\includegraphics[width=\textwidth]{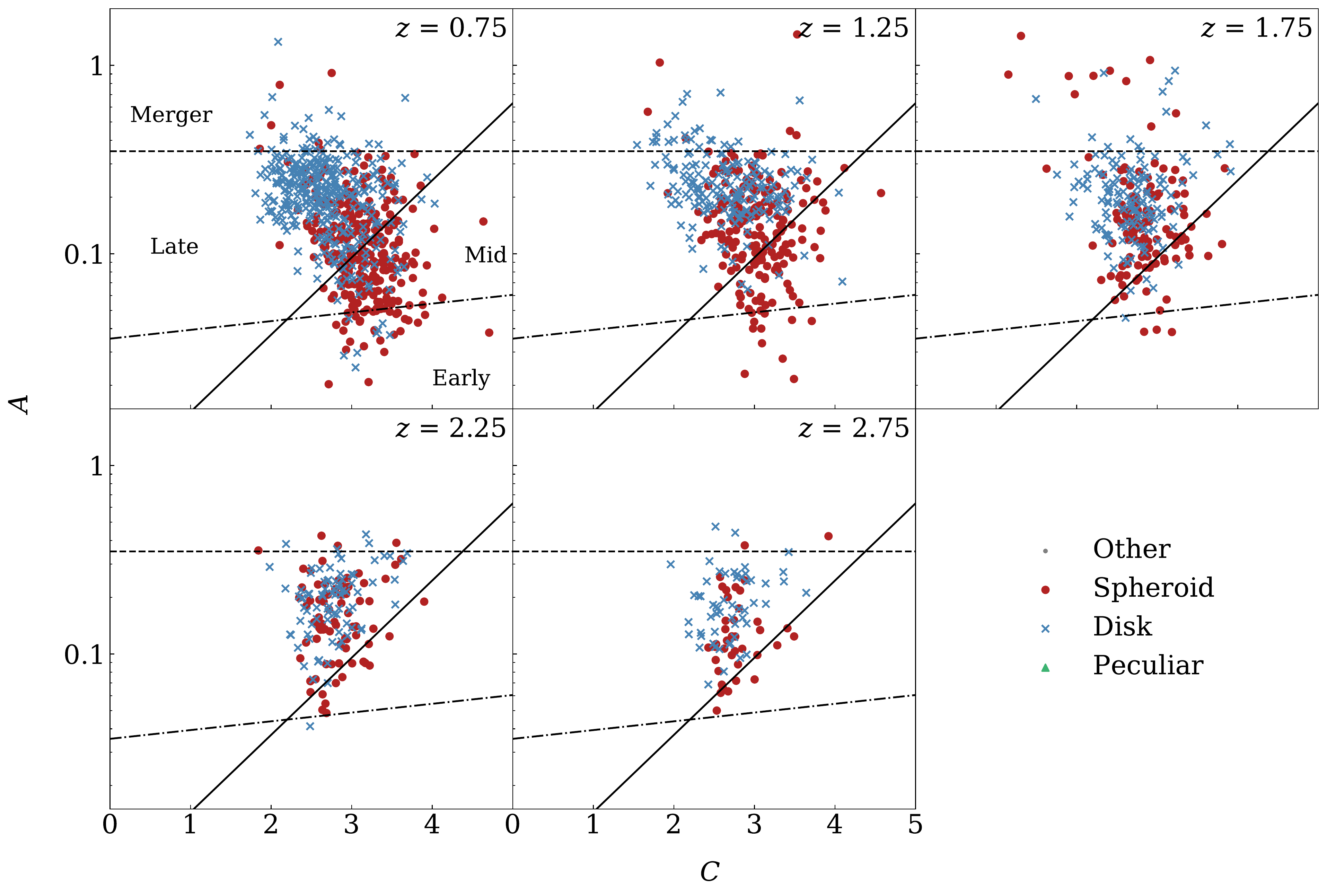}
\caption{Evolution of the concentration-asymmetry plane with redshift for galaxies that have been given either a bulge dominated flag (red circles) or a disk dominated flag (blue crosses). Galaxies that have neither flag are not plotted. As with Figure \ref{fig:vccva}, the lines are boundaries between galaxy types derived from \cite{bershady00}. The redshift indicated in the top right corner of each panel is the midpoint of the redshift bin whereby the upper and lower limits are 0.25 either side of this point.}
\label{fig:bdcva}
\end{figure*}

Within the catalog of \cite{kartaltepe15} every image has flags based on a number of interesting structural features present. Two of these flags indicated whether the galaxy appears to be bulge or disk dominated. As with the visual morphological classifications, we assume a galaxy is bulge or disk dominated if the fraction of classifications that designate it to be such is greater than 0.6, and the other classification is less than 0.6. We show the evolution with redshift of the concentration-asymmetry plane for those galaxies classified as being bulge or disk dominated in Figure \ref{fig:bdcva}. Disk dominated galaxies are shown as blue points and bulge dominated galaxies are shown as red points. As with Figure \ref{fig:vccva}, we also plot the boundaries between galaxy types. At higher redshifts there are fewer galaxies with either of the classifications in consideration here, likely due to greater noise and surface brightness dimming causing there to be less consensus among those classifying the images. This highlights the issues in classifying galaxy images and the need for a less subjective method of classifying galaxies.

Galaxies classified visually as disk-dominated predominantly lie in the late region of the concentration-asymmetry plane and the average values in the $0.5 < z < 1.0$ bin are $\langle A \rangle$ = 0.22 and $\langle C \rangle$ = 2.68. These values remain roughly constant across redshift and become $\langle A \rangle$ = 0.20 and $\langle C \rangle$ = 2.69 at $2.5 < z < 3.0$. At high redshifts, the bulge-dominated galaxies appear to lie in the late region but as redshift decreases, there are more galaxies in the intermediate and early regions of the plane. The average values of asymmetry and concentration of these bulge-dominated galaxies are $\langle A \rangle$ = 0.13 and $\langle C \rangle$ = 3.14 at $0.5 < z < 1.0$. The asymmetry value remains roughly constant to higher redshifts and becomes $\langle A \rangle$ = 0.14 at $2.5 < z < 3.0$ but the concentration decreases with redshift and becomes $\langle C \rangle$ = 2.80 at this redshift. 

\subsubsection{Clumpiness/Patchiness Matrix}

\begin{figure*}
\centering
\begin{tabular}{cc}
\subfloat{\includegraphics[width = .49\textwidth]{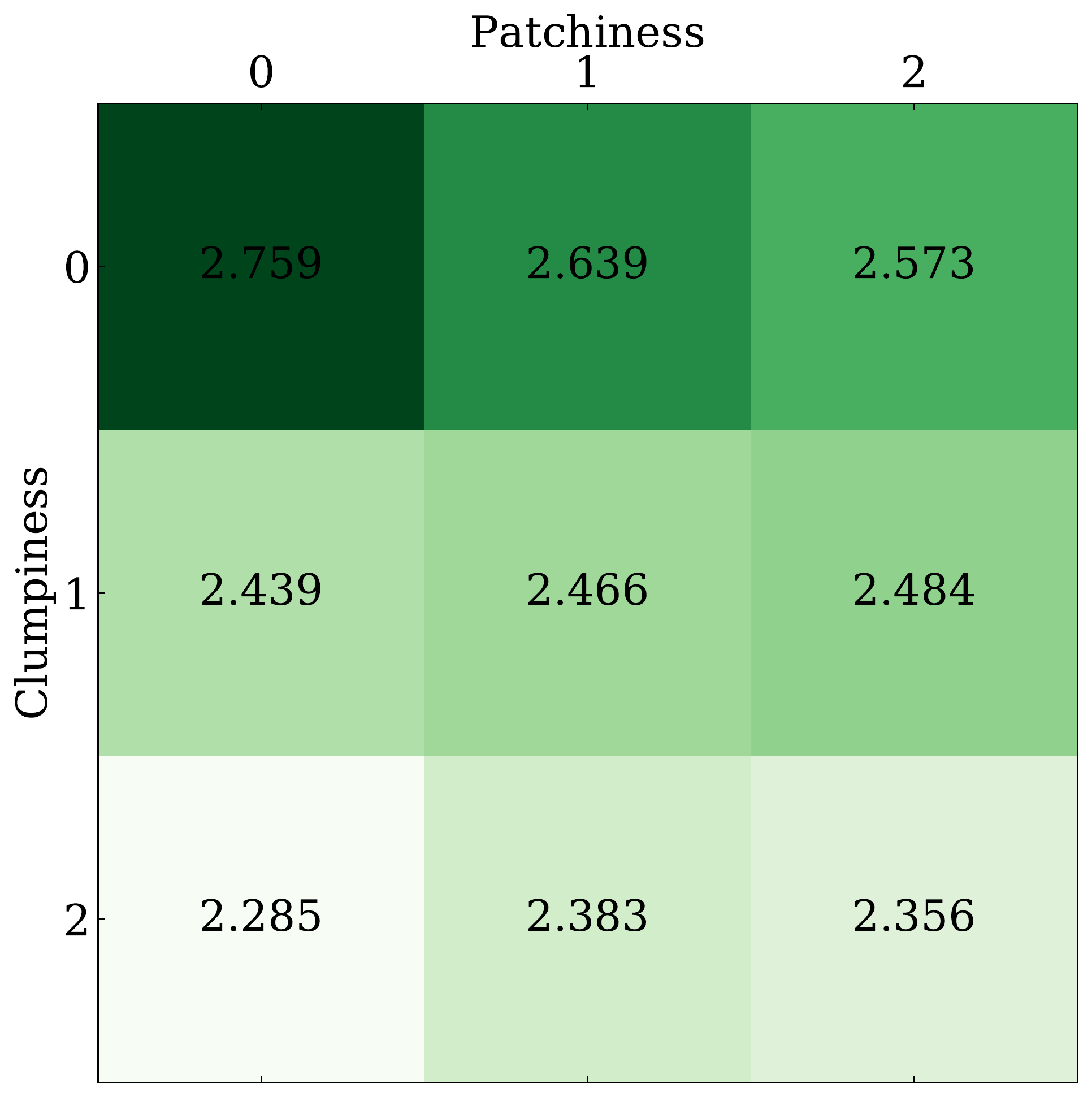}} &
\subfloat{\includegraphics[width = .49\textwidth]{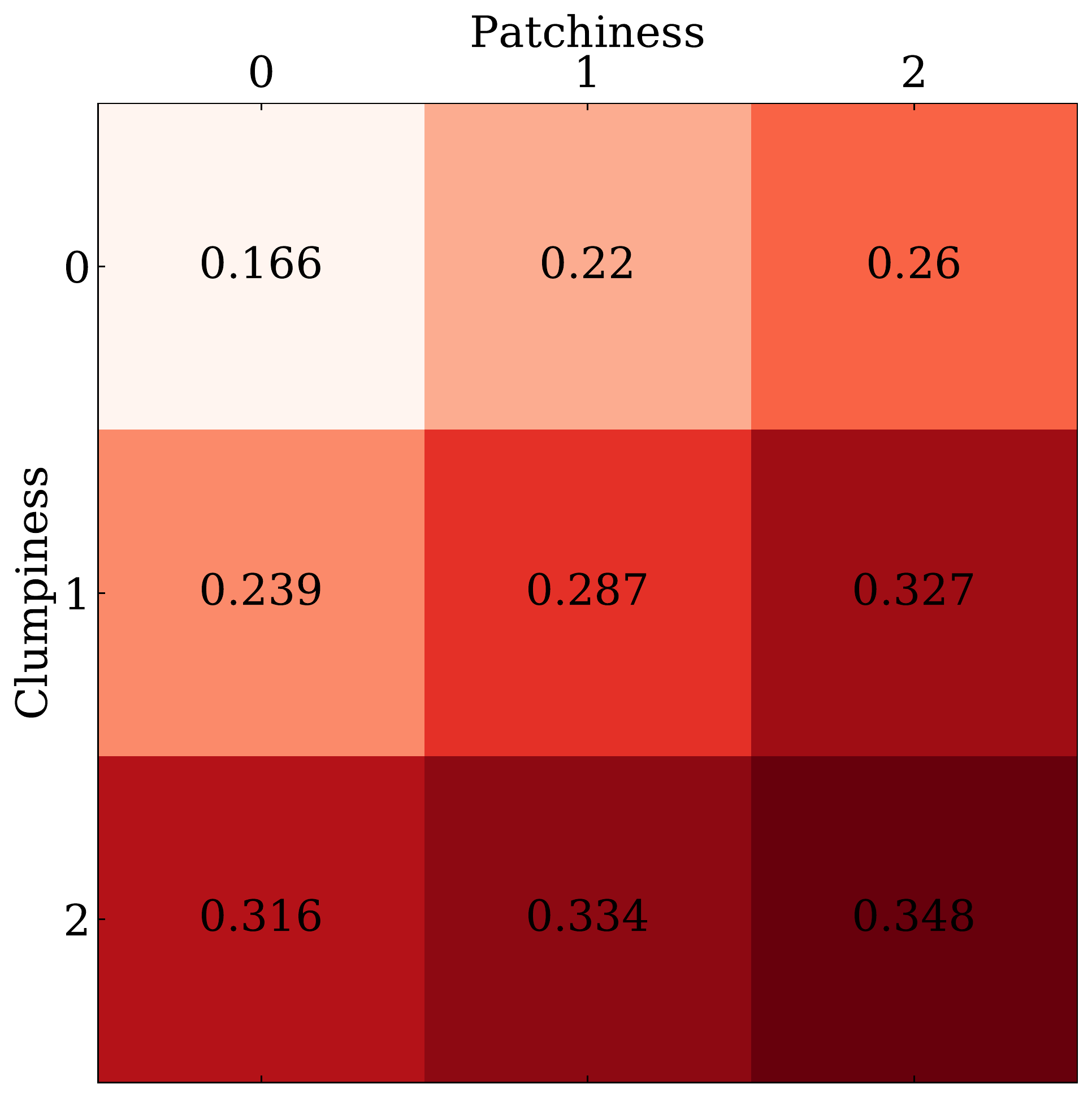}} \\
\subfloat{\includegraphics[width = .49\textwidth]{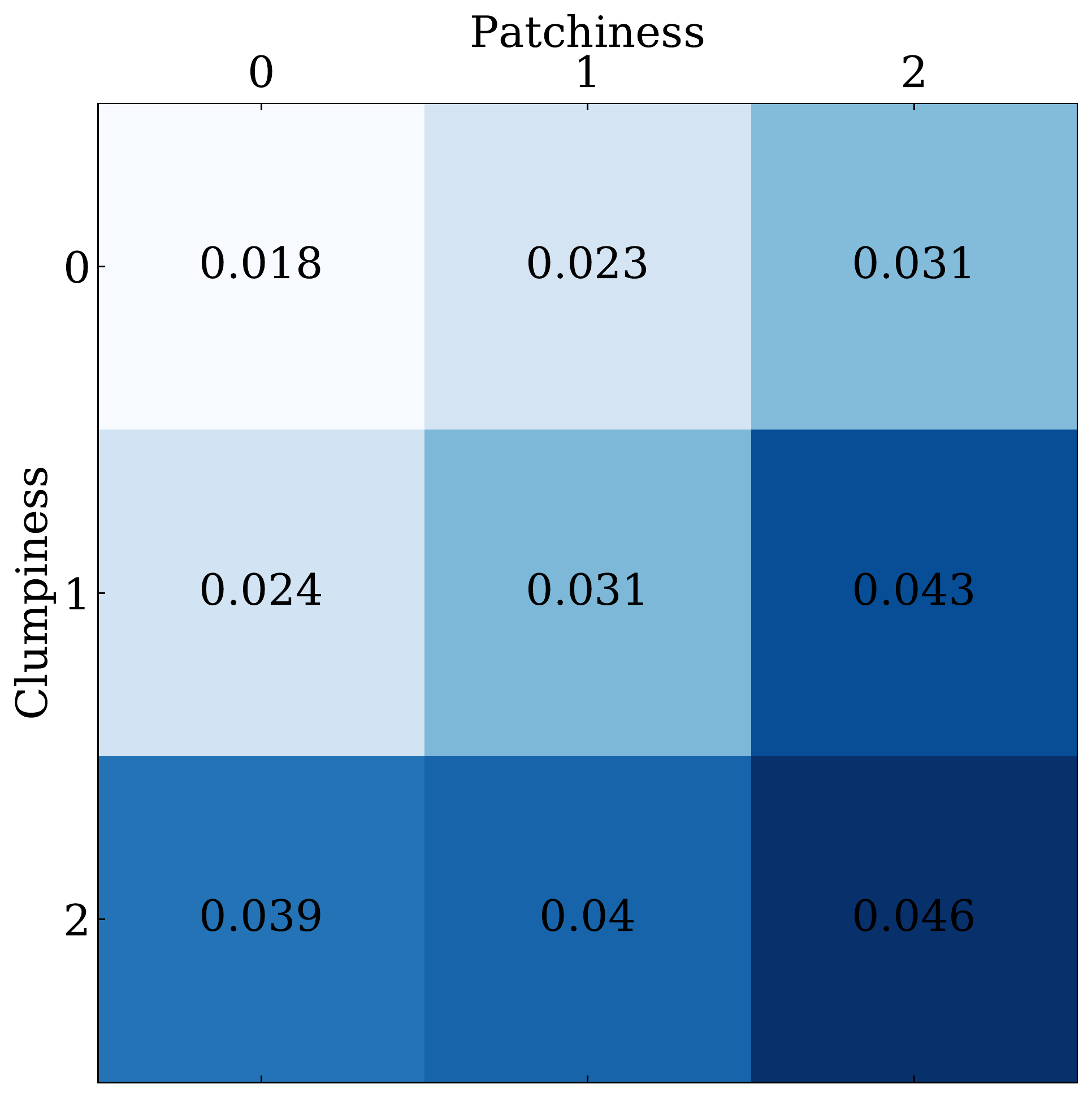}} &
\subfloat{\includegraphics[width = .49\textwidth]{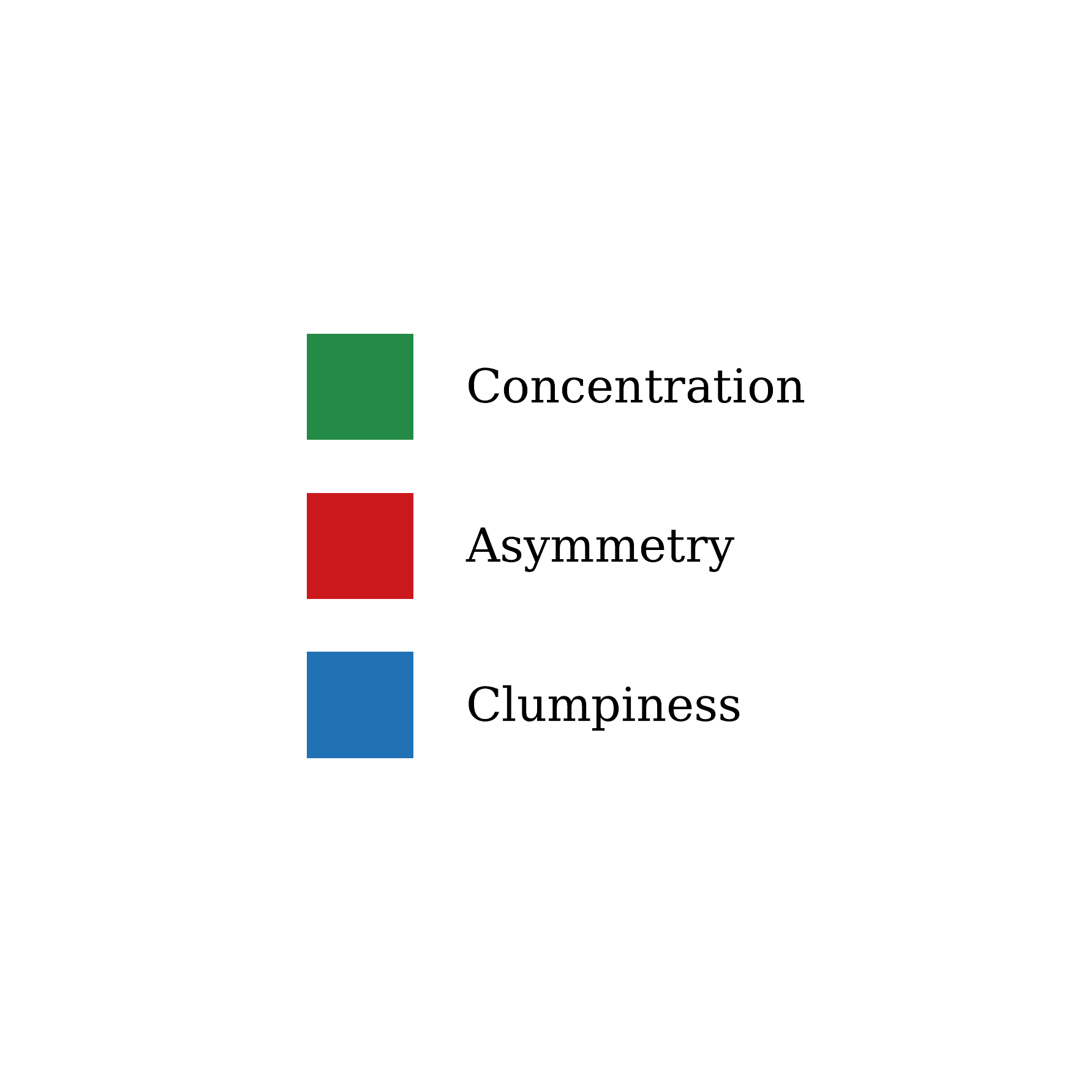}} \\
\end{tabular}
\caption{This figure shows the clumpiness (C) and patchiness (P) distribution visually vs. the CAS parameters. The values on the $x$- and $y$-axes are the visual estimates of clumpy galaxies, with high numbers denoting more visually looking clumpy systems. The clumpiness values on the $y$-axes are different to the clumpiness calculated by \textsc{Morfometryka} and are a purely visual score. Top left, green: mean concentration ($C$) values for each region of the clumpiness/patchiness matrix defined by \cite{kartaltepe15}. Top right, red: mean asymmetry ($A$) values for each region of the clumpiness/patchiness matrix. Bottom left, blue: mean clumpiness ($S$) values for each region of the clumpiness/patchiness matrix. In each panel, a darker colour indicates a larger value of the relevant CAS parameter. Each three by three grid represents the clumpiness/patchiness matrix whereby a higher value on each axis represents a higher clumpiness or patchiness. The visual clumpiness/patchiness criteria appear to correlate well with the measured $A$ and $S$ values whilst concentrated galaxies are not visually clumpy or patchy.}
\label{fig:matrix}
\end{figure*}

Along with visual classifications of galaxies, classifiers involved in the work of \cite{kartaltepe15} also assigned flags based on how clumpy or patchy the distribution of the light within the galaxies is. Clumps are defined to be concentrated knots of light and patches are defined to be more diffuse structures of light. There are 9 flags associated with this method of identifying features, with each flag denoting a different combination of the clumpiness (C) and patchiness (P). For example, a galaxy that exhibits no clumpiness and no patchiness will be labelled as 0C0P. A galaxy that appears to be extremely clumpy and patchy will be labelled as 2C2P. The levels of clumpiness and patchiness range between 0 and 2. As such we are able to organise the flags into a 3$\times$3 grid with each square of the grid denoting the level of clumpiness or patchiness. Note, this clumpiness is not the same as the clumpiness, $S$, described in \S\ref{sec:clumpiness} and is a purely visual descriptor. A score of 0 indicates no clumpiness or patchiness and a score of 2 indicates a large amount of clumpiness or patchiness.  In Figure \ref{fig:matrix}, we show this matrix of values and in each square of the grid, give the average concentration (top left, green), asymmetry (top right, red), and clumpiness (bottom left, blue). These average values are across all redshifts so any evolution in these parameters is not considered here. The lighter shades of colour denote a lower value of the CAS parameters while a darker shade denotes a higher value. In terms of concentration, the less clumpy or patchy a galaxy is, the higher its concentration on average. The opposite is true for the asymmetry and clumpiness; the more clumpy/patchy an image is, the higher the measured asymmetry and clumpiness. 

\subsection{CAS Parameter Evolution}

We now investigate with our data the evolution of the CAS parameters. We ultimately use image simulations to determine how these parameters change with the effects of resolution and noise removed. Ultimately the evolution of these parameters will lead to a physical understanding of the driving forces behind galaxy formation over the epoch $0.5 < z < 3$.

\subsubsection{Concentration Evolution}

\begin{figure}[!ht]
\includegraphics[width=0.475\textwidth]{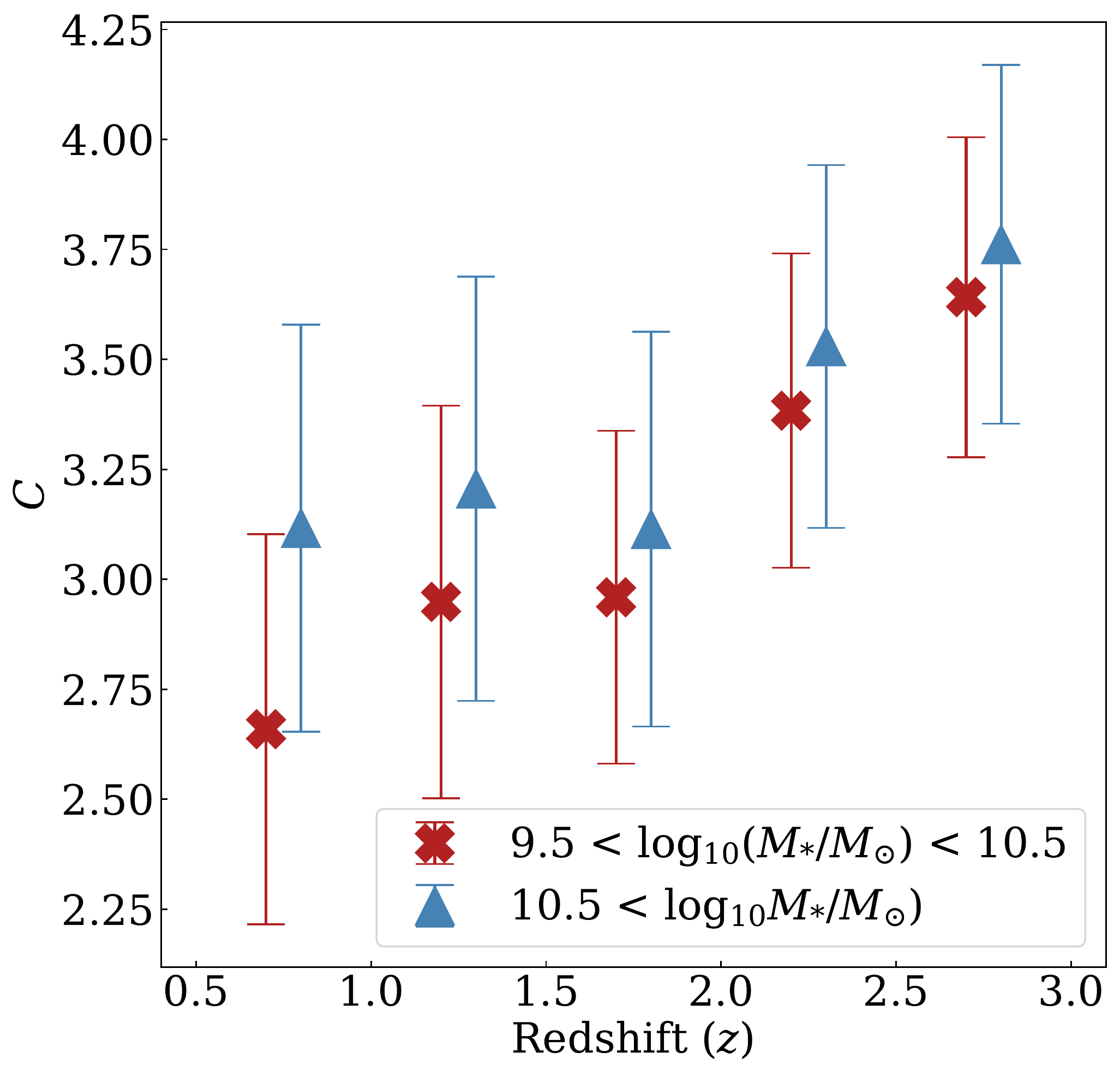}
\caption{Evolution of the mean corrected concentration for both mass bins. The lower mass bin is shown as a red cross and the higher mass bin is shown by the blue triangles. The error bars represent one standard deviation from the mean. On average, the higher mass bin has a greater concentration than the lower mass bin at all redshifts, however both bins show a decrease in concentration with cosmic time.}
\label{fig:conc_evol}
\end{figure}

The concentration of a galaxy tells us important information about how the light is distributed within a galaxy, relative to its centre. We show the evolution of the corrected concentration index defined in \S\ref{sec:conc} in Figure \ref{fig:conc_evol} for two different mass ranges across the redshift range $0.5 < z < 3$. The concentration values have been corrected using the method described in \S \ref{sec:corrections}. 

The first includes galaxies in the mass range $10^{9.5}M_{\odot} \leq M_{*} <  10^{10.5}M_{\odot}$ and the evolution of the mean concentration for this mass bin is shown as red crosses (see Figure \ref{fig:conc_evol}). The second mass bin includes galaxies that have a mass $M_{*} \geq 10^{10.5}M_{\odot}$. This mass bin is plotted as the blue triangles. For both, the error bars represent one standard deviation from the mean. The lower mass galaxies have a lower concentration than the higher mass galaxies, but both samples on average exhibit a decrease in concentration with time.   This is opposite to what we find when we examine the uncorrected concentration index evolution.  This therefore deserves some attention to try to understand the origin of this, and whether it is in fact a real effect.  

\subsubsection{Asymmetry Evolution}

\begin{figure}[!ht]
\includegraphics[width=0.475\textwidth]{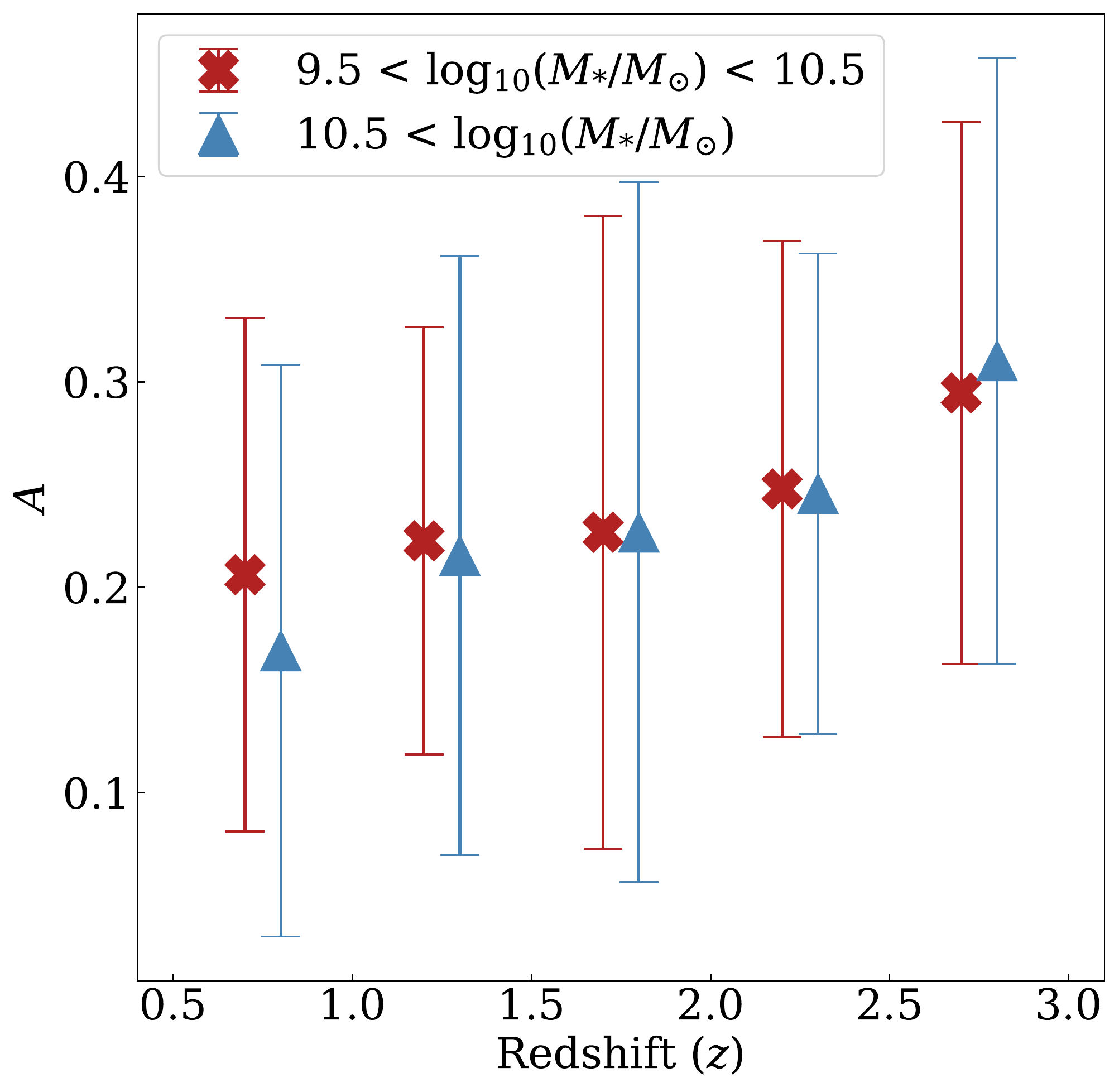}
\caption{Evolution of the mean corrected asymmetry with redshift. As with Figure \ref{fig:conc_evol}, the lower mass bin is shown as a red cross and the higher mass bin is shown by the blue triangles. The error bars represent one standard deviation from the mean. Both mass bins show a decrease in asymmetry with decreasing redshift.}
\label{fig:asym_evol}
\end{figure}

The asymmetry of a galaxy is a useful indicator as to whether a galaxy is undergoing any interactions or mergers with other galaxies \citep{conselice00b, conselice00a}. We explore its use to define mergers in the following section, but first we examine the evolution of the corrected asymmetry for our sample of galaxies in the redshift range $0.5 < z < 3$. Figure \ref{fig:asym_evol} shows this evolution for two different mass ranges. As in Figure \ref{fig:conc_evol}, the lower mass bin is shown as red crosses, the higher mass bin is shown as blue triangles, and the error bars represent one standard deviation from the mean. On average, the asymmetry decreases with cosmic time for both mass bins, with the higher mass bin exhibiting a steeper decrease. 

\subsubsection{Concentration-Asymmetry Plane}

\begin{figure*}[!ht]
\includegraphics[width=\textwidth]{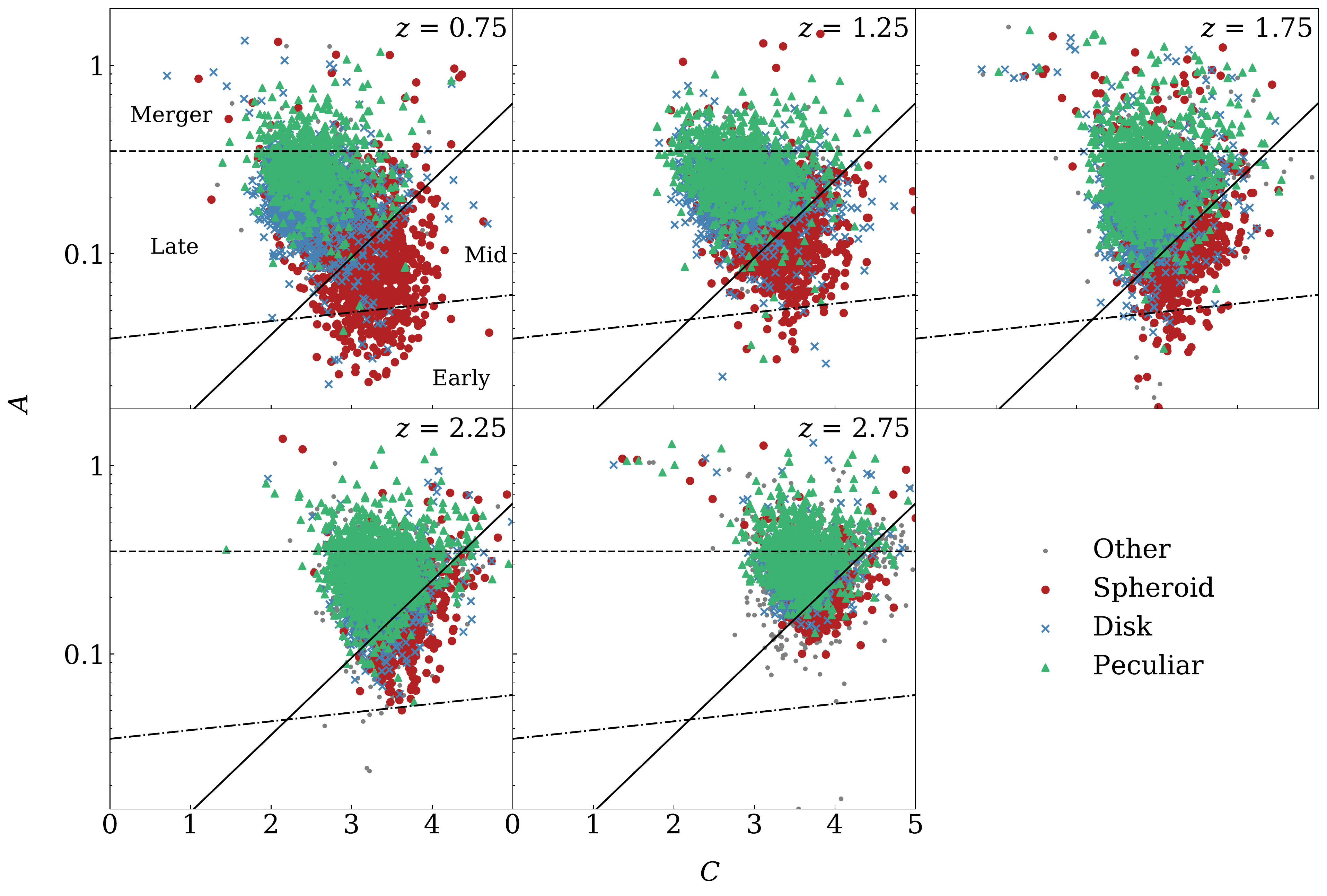}
\caption{Evolution of the corrected concentration-asymmetry plane with redshift. As with Figure \ref{fig:vccva}, the lines indicates the boundaries between galaxy types. Galaxies that lie above the dashed line are considered to be mergers. The dot-dashed line is the boundary between early- and intermediate-type galaxies and the solid line is the boundary between late- and intermediate-type galaxies. Spheroid galaxies are indicated by red circles, disks by blue crosses, and peculiar galaxies by green triangles. Galaxies that do not fall into any of the three categories are shown as small grey circles. The redshift indicated in the top right corner of each panel is the midpoint of the redshift bin whereby the upper and lower limits are 0.25 either side of this point.}
\label{fig:cva}
\end{figure*}

In Figure \ref{fig:cva}, we show the evolution of the concentration-asymmetry plane for the corrected values with redshift. As with Figures \ref{fig:vccva} and \ref{fig:bdcva}, we show the boundaries between the galaxy types. We also plot the three visual classifications for each galaxy, along with the position of any galaxy that does not fit these three categories within this plane. Spheroid galaxies are shown as red circles, disks as blue crosses, and peculiar galaxies as green triangles. Any other galaxies are shown as small grey circles. The spread in distribution of both concentration and asymmetry increases at lower redshifts, leading to the evolution of the average of both these parameters as seen in Figures \ref{fig:conc_evol} and \ref{fig:asym_evol}. 

There are few galaxies that lie within the early region of the concentration-asymmetry plane and those that are are typically spheroid galaxies. This suggests that the galaxies within our sample are more asymmetric than is typical for spheroids. Spheroid galaxies that do not lie in this early region are mostly within the intermediate region. Within the lowest redshift bin ($0.5 < z < 1.0$), spheroids have a mean asymmetry of $\langle A \rangle$ = 0.14 and a mean concentration of $\langle C \rangle$ = 3.03. At the highest redshift ($2.5 < z < 3.0$), these values change to $\langle A \rangle$ = 0.25 and $\langle C \rangle$ = 3.77 with a steady increase in both $A$ and $C$ at intermediate redshifts. Disk galaxies are primarily located within the late region, as to be expected. The mean values are $\langle A \rangle$ = 0.21 and $\langle C \rangle$ = 2.52 at $0.5 < z < 1.0$ and change to $\langle A \rangle$ = 0.28 and $\langle C \rangle$ = 3.62 at $2.5 < z < 3.0$. The change in $C$ is a steady increase however asymmetry remains at an approximately constant value until the highest redshift bin. Peculiar galaxies lie around the merger limit of $A >$ 0.35 with 29.20 $\pm$ 0.01\% of galaxies classified as being peculiar lying within this merger region.   If these corrections are right, we can see that there is a diversity already at $2.5 < z < 3.0$ in the galaxy population and that galaxies of all types are more asymmetric and more concentrated at higher redshifts.

\subsection{Merger Fractions and Rates}

This section of the paper describes our results of the merger history of galaxies within the five CANDELS fields using a CAS approach. The outline of this section is as follows: first we give a background description of the merger history of galaxies, include definitions of the merger process.  We then describe in some detail how to measure the timescale for mergers and how these can then be used for measuring the merger rate of galaxies.  Finally, from the merger rates we are able to say how many mergers these galaxies undergo on average and how much mass is added to galaxies through this process. 

\subsubsection{Merger Fractions} \label{sec:merge_frac}

In this subsection we investigate the observational quantity that we can obtain directly from the data. This is the merger fraction, which in this paper we use the CAS parameters described in \S \ref{sec:morf} to define. We later compare the merger fractions and the merger rates we derive using merger timescales to other measurements of the merger history using machine learning methods \citep{ferreira20} and galaxies in pairs \citep{duncan19}. First we define what we mean by a merger fraction - that is, the fraction of galaxies within some galaxy sample which is undergoing a major merger.

\begin{figure*}[!ht]
\includegraphics[width=\textwidth]{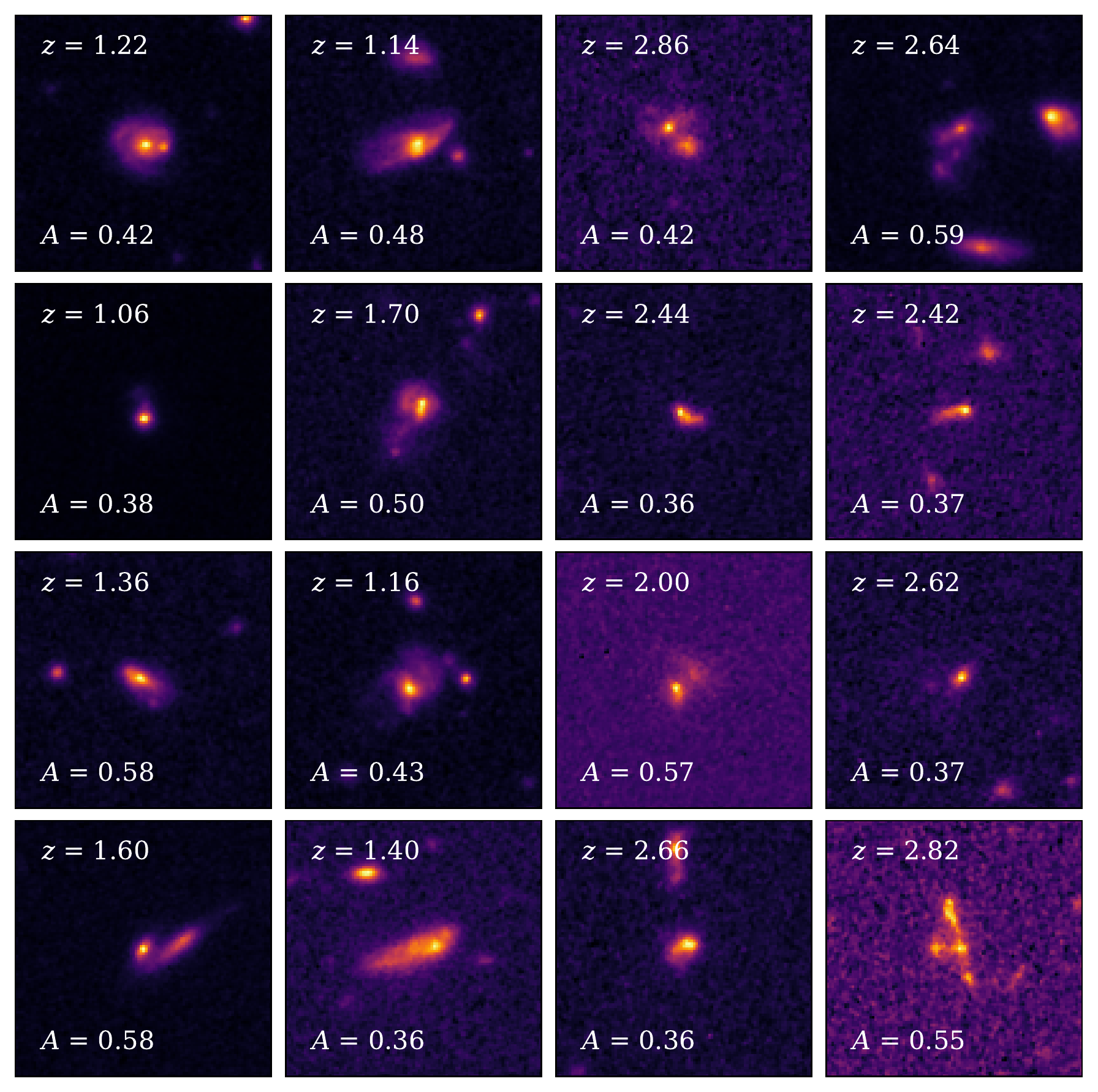}
\caption{Examples of galaxies within our sample that are classified as a merger using the criteria given in Equation \ref{eq:mergers}. Lower redshift ($1 < z < 2$) galaxies are shown in the left two columns and higher redshift ($2 < z < 3$) galaxies are shown in the right two columns. The redshift and asymmetry are given for each example.}
\label{fig:merger_ex}
\end{figure*}

In our sample, a merger is defined within the CAS system as a galaxy that satisfies the following criteria:

\begin{equation}
    (A > 0.35)\ \&\ (A > S)
    \label{eq:mergers}
\end{equation}

\noindent where both $A$ and $S$ are the asymmetry and clumpiness values corrected for redshift effects. This method predominantly identifies only major mergers where the ratio of the stellar masses of the progenitors is at least 1:4 \citep{conselice03, conselice06b, lotz08}. $H$-band images of examples of galaxies we consider to be a merger, based on this system, are shown in Figure \ref{fig:merger_ex}. The left two columns show lower redshift ($1 < z < 2$) galaxies, and the right two columns show higher redshift ($2 < z < 3$) galaxies. The redshift and asymmetry of each example are indicated on each postage stamp.

We calculate the merger fraction using this equation for each redshift bin, the evolution of which is shown in the left panel of Figure \ref{fig:merge_evol}. The results for the low mass bin ($10^{9.5}M_{\odot} \leq M_{*} <  10^{10.5}M_{\odot}$) are shown as red crosses and the results for the higher mass bin ($M_{*} \geq 10^{9.5}M_{\odot}$) are shown as blue triangles. We fit a power law to the merger fraction evolution given by the form:

\begin{equation}
    f_m(z) = f_0(1+z)^{\beta}
\end{equation}

\noindent by applying a least squares fit to our data and setting $f_0$ to a fixed value of the merger fraction at $z = 0$. We take this $z \sim 0$ value to be 0.0193 $\pm$ 0.0028 from \cite{casteels14}. This point is shown as a grey inverted triangle. We also include both mass bins in this fit. We find that the merger fraction, $f_{m}$ evolves as 

\begin{equation}
    f_{m} = 0.0193 \pm 0.004 (1+z)^{1.87 \pm 0.04}.
\end{equation}

\noindent This fit is shown in the left panel of Figure \ref{fig:merge_evol} as a solid black line and one standard deviation from this fit is indicated by the grey shaded area. Also shown are the results of \cite{duncan19} (yellow circles) who identify mergers using pair statistics and fit a power law of the form $(1+z)^{1.775^{+0.205}_{-0.196}}$. We also compare to \cite{ferreira20} (green diamonds) who identify mergers using a convolutional neural network and find that the merger fraction evolves as $(1+z)^{2.82 \pm 0.46}$. Our merger fractions lie between these two examples. The merger fractions and their errors are given in Columns 2 and 4 of Table \ref{tab:merge_frac} for both the lower and higher mass bins respectively. 

\subsubsection{IR Luminous Galaxies}

One of the things that we need to consider in our approach is that some galaxies are invisible to optical/NIR light and are only detected in the far-IR, sub-mm or radio \citep[e.g.,][]{wang19}.  These are sometimes called $H$-band drop-outs and are of interest as they may in fact be galaxies that are massive, within our limits, but are not considered part of the merger fraction as they would not be added into our $H$-band selected sample.

A simple argument can be made to show that this concern is insignificant for our merger fractions.  The study of \cite{wang19} found 39 galaxies across the same fields we study, which are not detected in the $H$-band, and thus would likely not be within our sample. \cite{wang19} claim that these galaxies are typically massive ones at $z > 2$, with a median stellar mass of $10^{10.6} M_{\odot}$. Let us examine the results of assuming that all of these galaxies are mergers. 

For our CANDELS sample, we have 6273 massive galaxies at $z > 2$, whilst the merger fraction we measure is $f_{m}=0.183 \pm 0.005$ for these galaxies. This gives in total 1135 mergers within our sample.  Then if we consider all 39 \cite{wang19} galaxies to be mergers at $z > 2$ and within the mass range we are looking at, then the new merger fraction with the \cite{wang19} drop-outs included would be (1135+39)/(6273+39) = 0.186.  This gives a very small merger difference of $\sim 0.003$. The Poisson error on our measurements is also 0.005, at the same level. If we consider that these galaxies have the same merger fraction as the bulk of our systems, at $f_{m} = 0.183$, then the merger fraction would be: $f_{m} = 0.183$, a difference of $< 0.001$ - almost unchanged and vastly lower than our counting errors.

Regardless, if we consider these FIR galaxies as mergers or that have a merger fraction higher than our measured one, it would only increase our values extremely slightly and well within our current measurement errors. 

\subsubsection{Merger Timescales from IllustrisTNG}\label{subsec:timescales_tng}

In this section we investigate the merger timescales for a sample of galaxies in the IllustrisTNG 300-1 simulation.  The process we use to generate mock images is outlined in \S\ref{subsec:tng} and here we use only \textit{HST}-matched mocks. We measure the observability timescales with a process similar to the one developed in \cite{lotz08} and \cite{Nevin19}. The main difference here is that the galaxies used in this evaluation are generated as a result of a cosmological simulation instead of an isolated galaxy-galaxy merger simulation. Our sample is limited to massive $M_{*} > 10^{10} M_\odot$ major-mergers. We briefly outline our steps to measure the timescales below.

First, we follow each galaxy in the simulation for a variety of snapshots spanning $\pm \ 2 \ \rm Gyr$ around each merger event (limited to $0 < z < 2$). The snapshot where the merger event happens is defined as the the central snapshot, $S_c$. This information is extracted from the merger trees in the simulation and is originally defined by the SubLink friends-of-friends algorithm \citep{Rodriguez-Gomez2015}. The snapshots after $S_c$ are considered to be after the merger event, while the snapshot previous to $S_c$ are classified as before the merger event. This is the only distinction done here for particular merger stages, which is fundamentally different to the stages assigned in \cite{lotz08}. 

Secondly, we select only the broadband images that correspond to the rest-frame optical at those redshifts. For these images, we measure the asymmetry using \textsc{Morfometryka}. Thus, for each source simulated in \S \ref{subsec:tng} we have $A$ for each snapshot in four different orientations. By averaging $A$ by the viewing angle, we find where the asymmetries of each individual galaxy falls below the $A > 0.35$ threshold for both sides of the merger event. 

\begin{figure}[!ht]
\includegraphics[width=0.475\textwidth]{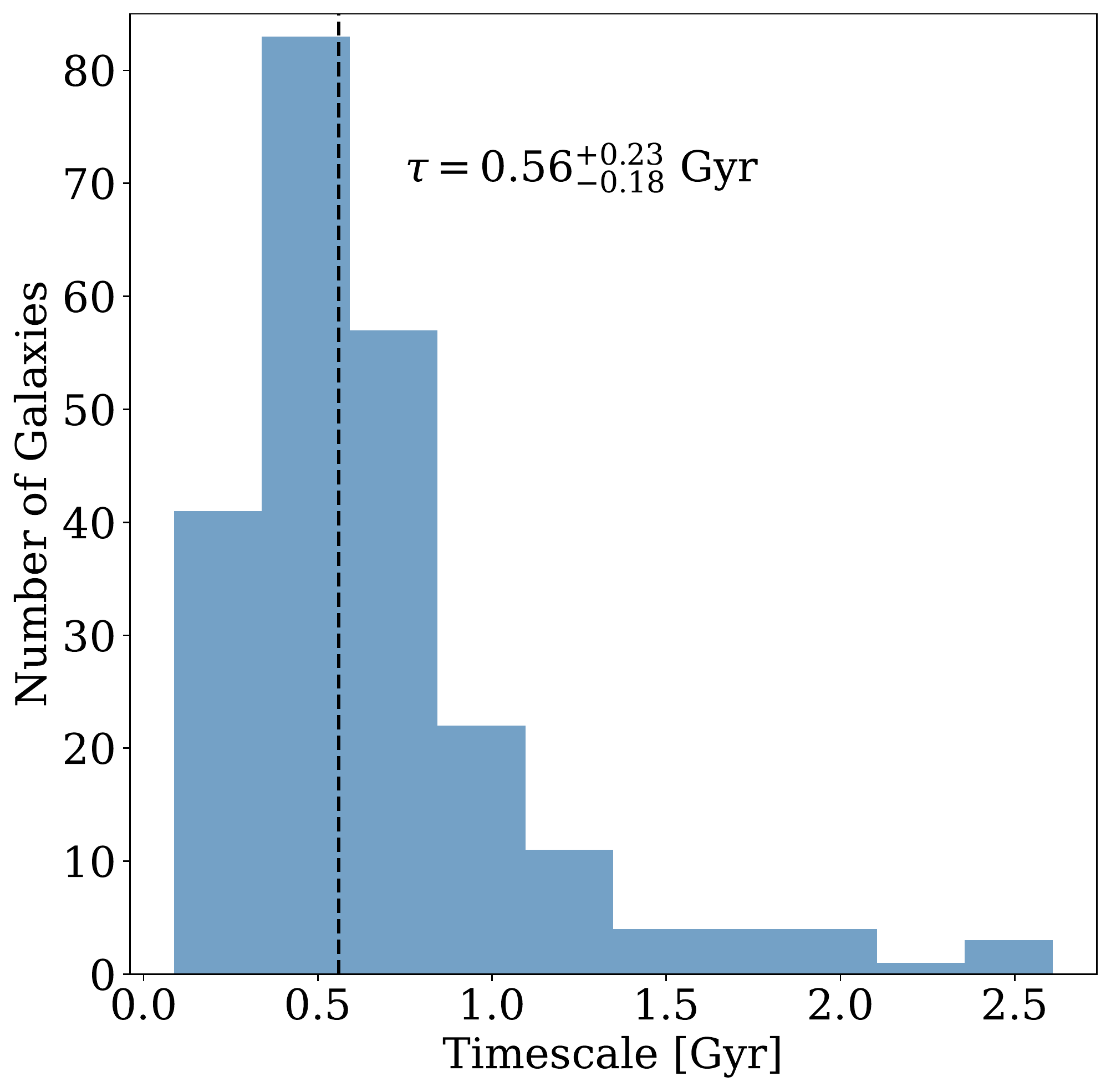}
\caption{Distribution of the timescales for the total merger events within the IllustrisTNG sample. The dashed line indicates the median timescale used to determine the merger rates.}
\label{fig:timescale}
\end{figure}

Finally, by comparing the time difference between the redshift where the asymmetry threshold is no longer valid and the redshift of the central snapshot, we estimate an observability timescale for the asymmetry, both for the post-merger stage and the before merger stage, and, if combined, for the total merger event. By doing this for all our sample, we have a statistical estimation of the merger timescales for the asymmetry in IllustrisTNG, as shown in Figure \ref{fig:timescale}. We find a $\tau = 0.56$, lower than the values reported in \cite{lotz08}, but higher than the ones found in \cite{Nevin19} for major-mergers. We find an asymmetric distribution in the timescales that is broad but not deep; there is an asymmetric tail at higher timescales. The mean timescale of this distribution is $\sim$20\% greater than the median at 0.67 Gyr so the tail does not significantly impact the timescale used. We consider the tail in our error in the timescale by finding the differences between the median of the full sample of simulations and the median of the sample both above and below 0.56 Gyr. From this, we yield a timescale of $0.56^{+0.23}_{-0.18}$ Gyr. We find that this timescale does not vary with redshift unlike in the case of pair statistics \citep{snyder17}. 

\subsubsection{Merger Rates}

Merger fractions are a solely observational quantity and can tell us little about the evolution of galaxy formation and evolution. The merger rate (time between mergers per galaxy) on the other hand is a fundamental parameter in galaxy evolution. We are able to convert from a merger fraction to a the merger rate ($\mathcal{R}$) at each redshift using

\begin{equation}
        \mathcal{R} = \frac{f_{m}}{\tau}
        \label{eq:merger_rate}
\end{equation}

\noindent where $f_m$ is the merger fraction and $\tau$ is the merger timescale. 

\begin{figure*}
\centering
\begin{tabular}{cc}
\subfloat{\includegraphics[width = .49\textwidth]{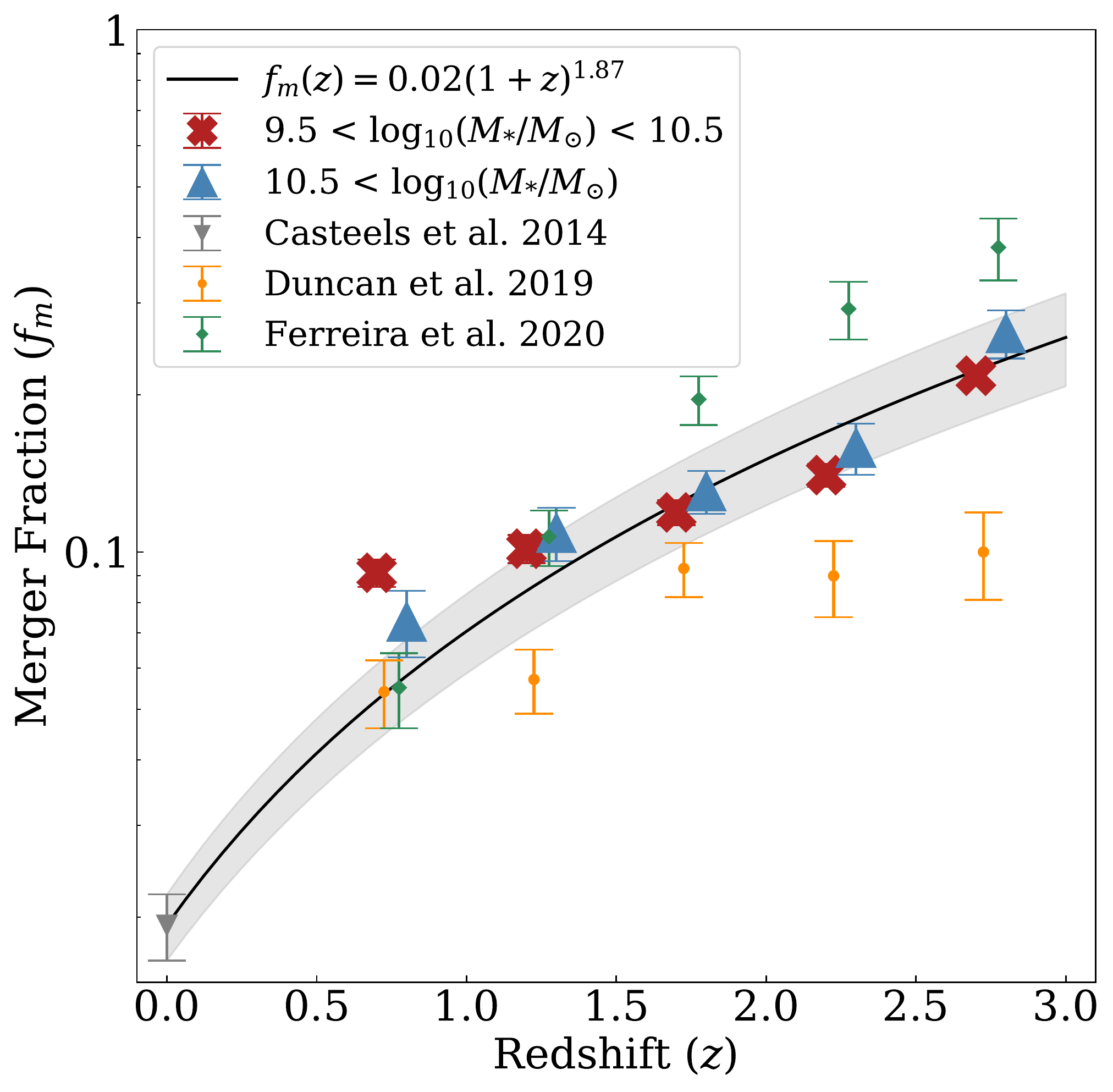}} &
\subfloat{\includegraphics[width = .49\textwidth]{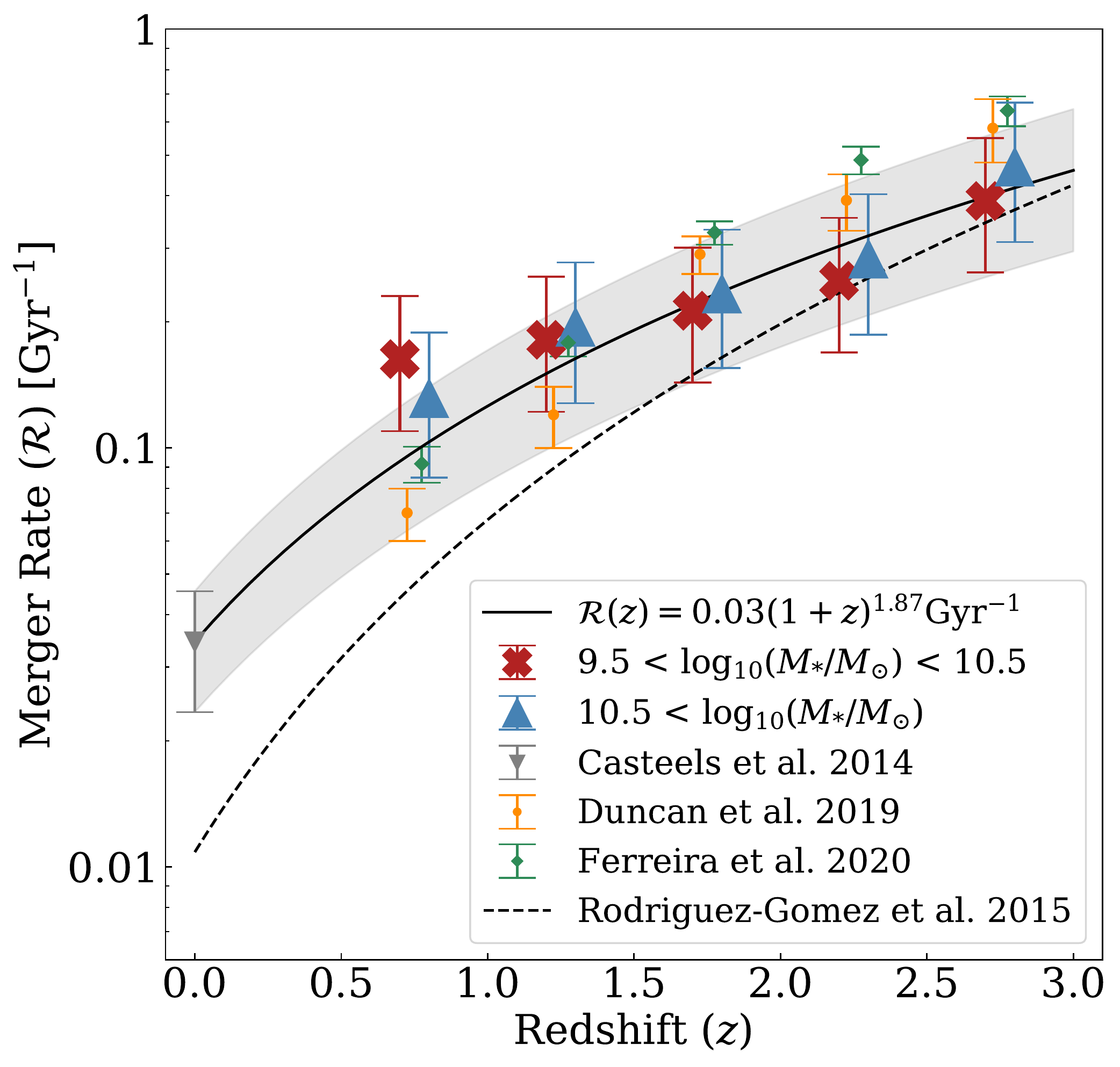}} \\
\end{tabular}
\caption{Left: evolution of the merger fraction. Right: evolution of the merger rate. The lower mass bin from this work is shown as a red cross and the higher mass bin from this work is shown as a blue triangle. The fit to both these points and a known local value from \cite{casteels14} (grey inverted triangle) is shown as a solid black line with the grey shaded area indicating one standard deviation from the fit. We find a fit of the form $f_{m} = 0.0193 \pm 0.004(1+z)^{1.87 \pm 0.04}$ for the merger fractions. For the merger rates, we find a fit of the form $\mathcal{R}(z) = 0.03 \pm 0.02(1+z)^{1.87 \pm 0.04}$ Gyr$^{-1}$. We compare our results to \cite{duncan19} (yellow circles) and \cite{ferreira20} (green diamonds). We also compare the merger rate to that found by \cite{Rodriguez-Gomez2015} for the Illustris simulation. This fit is given as a dashed line.}
\label{fig:merge_evol}
\end{figure*}

We calculate the merger rate evolution for our sample of galaxies using a timescale of $\tau = 0.56^{+0.23}_{-0.18}$ Gyr found using mergers within the IllustrisTNG simulations as described above. The merger rate evolution is shown in the right panel of Figure \ref{fig:merge_evol}. As with the merger fraction, we fit a power law to both the low-mass (red crosses) and high-mass (blue triangles) data of the form 

\begin{equation}
    \mathcal{R}(z) = \mathcal{R}_0(1+z)^{\beta} \  \rm  Gyr^{-1}
\end{equation}

\noindent where $\mathcal{R}_0$ is the local merger rate calculated using the local merger fraction, $f_0$, from \cite{casteels14} and the merger timescale we find using the IllustrisTNG simulations. This local merger rate is shown as a grey inverted triangle. We find that the merger rate evolves as:

\begin{equation}
    \mathcal{R}(z) = 0.03 \pm 0.02(1+z)^{1.87 \pm 0.04} \  \rm  Gyr^{-1}.
    \label{eq:mergerrate}
\end{equation}

This fit is shown by a solid black line and one standard deviation from this fit is indicated by the grey shaded area in the right panel of Figure \ref{fig:merge_evol}. As with the merger fractions, we also show the results of \cite{duncan19} and \cite{ferreira20}. Due to the different methods used to identify mergers, different timescales must be used; \cite{ferreira20} use $\tau = 0.6$ Gyr and \cite{duncan19} use a timescale that varies with redshift such that $\tau$ = 0.17 Gyr at $2.5 < z < 3.0$ and $\tau$ = 0.77 Gyr at $0.5 < z < 1.0$ in order to calculate their merger rates. 

Despite using a timescale that differs from other merger identification methods, our results are largely consistent with these previous results. Our merger rates and errors for the low and high mass bins are given in Columns 3 and 5 of Table \ref{tab:merge_frac} respectively. We also compare our results to those found by \cite{Rodriguez-Gomez2015} for the Illustris simulation. This fit is shown by a dashed line here. The fit, given in Table 1 of \cite{Rodriguez-Gomez2015}, is a complex function of redshift, the descendent galaxy stellar mass $M_{*}$, and the progenitor stellar mass ratio $\mu_*$. We show the results for a stellar mass of $1 \times 10^{10}M_{\odot}$ and a stellar mass ratio of 1/4 (consistent with major mergers, the type of mergers we are able to probe with our asymmetry selection). The Illustris fit is steeper than the results we find in this paper and can be approximated as a power law of the form $\sim (1+z)^{2.7}$. These simulations also predict a smaller merger rate, particularly at the lowest redshifts, with the local merger rate predicted to be $\sim$3 times smaller than the value we use here. 

\begin{table*}
\centering
\caption{Merger fractions and rates for each redshift bin and each mass bin.}
  \begin{tabular}{ccccc}
  \hline 
  \hline
   & \multicolumn{2}{c}{$10^{9.5}M_{\odot} \leq M_{*} <  10^{10.5}M_{\odot}$} & \multicolumn{2}{c}{$10^{10.5}M_{\odot}\leq M_{*}$} \\
  $z$ & $f_{m}$ & $\mathcal{R}$ (Gyr$^{-1}$) & $f_{m}$ & $\mathcal{R}$ (Gyr$^{-1}$)\\
  \hline
  0.75 & 0.091 $\pm$ 0.006 & $0.163^{+0.068}_{-0.053}$ & 0.074 $\pm$ 0.011 & $0.131^{+0.057}_{-0.046}$ \\
  1.25 & 0.102 $\pm$ 0.006 & $0.181^{+0.075}_{-0.059}$ & 0.108 $\pm$ 0.013 & $0.194^{+0.083}_{-0.066}$ \\
  1.75 & 0.119 $\pm$ 0.006 & $0.213^{+0.088}_{-0.069}$ & 0.131 $\pm$ 0.012 & $0.233^{+0.098}_{-0.078}$ \\
  2.25 & 0.140 $\pm$ 0.007 & $0.251^{+0.104}_{-0.082}$ & 0.158 $\pm$ 0.018 & $0.283^{+0.120}_{-0.096}$ \\
  2.75 & 0.217 $\pm$ 0.009 & $0.388^{+0.160}_{-0.126}$ & 0.263 $\pm$ 0.028 & $0.469^{+0.199}_{-0.158}$ \\
  \label{tab:merge_frac}
  \end{tabular}

\end{table*}

\subsubsection{Number of Mergers Since \texorpdfstring{$z \sim 3$}{}}

From the merger rate we are able to calculate the number of mergers a galaxy at $0 < z < 3$ undergoes by integrating the inverse of the characteristic time between mergers, $\Gamma(z)$. First, we must convert the merger fraction, $f_m$, to the galaxy merger fraction, $f_{gm}$, which gives the number of galaxies merging as opposed to the number of mergers \citep{bluck09}. The galaxy merger fraction is given by

\begin{equation}
    f_{gm} = \frac{2 \times f_m}{1+f_m}.
\end{equation}

\noindent We can then calculate $\Gamma$, which is essentially the time in between mergers, using this galaxy merger fraction:

\begin{equation}
    \Gamma = \frac{\tau}{f_{gm}}
\end{equation}

\noindent where $\tau$ is the merger timescale. We are able to fit a power law to $\Gamma$ of the form $\Gamma(z) = \Gamma_0(1+z)^{-m}$ Gyr and find that $\Gamma_0 = 14.4 \pm 0.7$ Gyr and $m = 1.96 \pm 0.13$. We calculate the number of mergers a given galaxy will undergo between redshifts $z_1$ and $z_2$, $N_{\textup{m}}$, using

\begin{equation}
    N_{\mathrm{m}} = \int^{t_2}_{t_1} \frac{1}{\Gamma(z)} dt = \int^{z_2}_{z_1} \frac{1}{\Gamma(z)} \frac{t_H}{1 + z} \frac{dz}{E(z)}
\end{equation}

\noindent where $t_H$ is the Hubble time, and $E(z) = [\Omega_m(1+z)^3 + \Omega_{\textup{k}}(1+z)^2 + \Omega_{\Lambda}]^{\frac{1}{2}} = \frac{H(z)}{H_0}$. We assume $\Omega_m = 0.3$, $\Omega_{\textup{k}} = 0$, and $\Omega_{\Lambda} = 0.7$. 

\begin{figure}[!ht]
\includegraphics[width=0.475\textwidth]{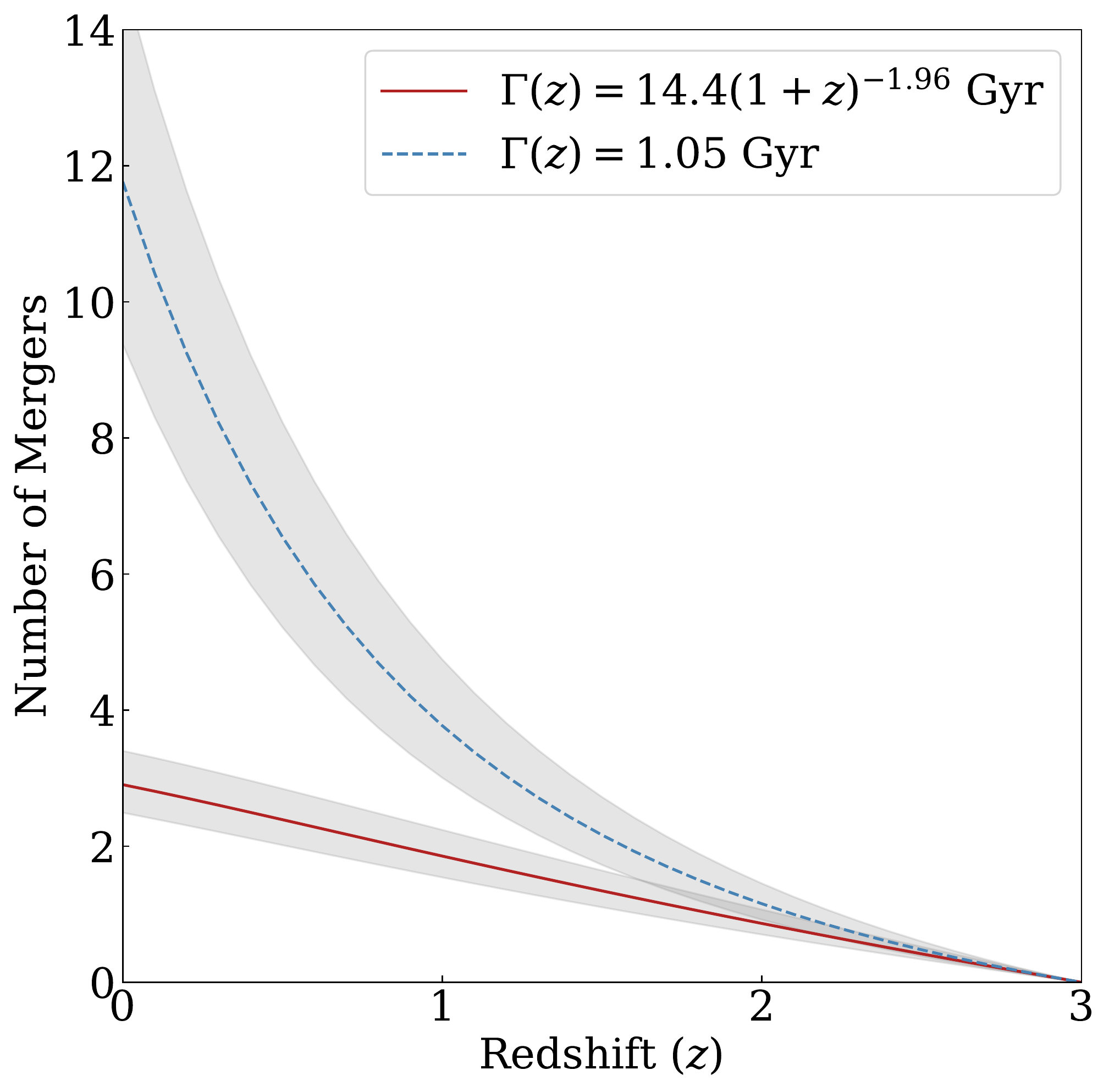}
\caption{The total number of mergers since $z \sim 3$, as a function of redshift. The solid red line shows this quantity for $\Gamma = 14.4 \pm 0.7(1+z)^{-1.96 \pm 0.13}$ Gyr. We find that galaxies with $M_{*} > 10^{9.5}M_{\odot}$ undergo $\sim$ 3 mergers on average since $z \sim 3$. For comparison, also shown is the evolution of the number of mergers for a constant merger rate of $\Gamma = 1.05$ Gyr, which is equal to the merger rate at $z = 3$. This is indicated by a blue dashed line. The grey shaded regions indicate the error range on the number of mergers.}
\label{fig:no_m}
\end{figure}

The resulting number of mergers undergone between redshifts $z \sim 3$ and $z \sim 0$ using the time between mergers of $\Gamma(z) = 14.4 \pm 0.7(1+z)^{-1.96 \pm 0.13}$ Gyr is shown in Figure \ref{fig:no_m} as a solid red line. We estimate that a galaxy with mass $M_{*} > 10^{9.5} M_{\odot}$ will undergo $2.90_{-0.41}^{+0.50}$ major mergers on average between $z \sim 3$ and $z \sim 0$, with $1.87_{-0.31}^{+0.39}$ mergers occurring before $z \sim 1$. This result lies between some previous results; \cite{bluck09} who find that massive ($M_{*} > 10^{11} M_{\odot}$) galaxies undergo $1.7 \pm 0.5$ major mergers with a timescale of $\tau = 0.4$ Gyr from $z \sim 3$ to $z \sim 0$. \cite{conselice06b} find that galaxies with $M_{*} > 10^{10} M_{\odot}$ undergo $4.4^{+1.6}_{-0.9}$ major mergers at $z > 1$ but undergo less than a single merger at $z < 1$. For comparison, we also plot the evolution of the number of mergers for a constant merger rate where we set $\Gamma(z)$ to be that at $z = 3$ to show how many mergers would occur without the observed decrease in merger rate. This is shown as a blue dashed line. The grey shaded areas on both lines indicate the error on the number of mergers. 

We are then able to determine how much mass is added to a galaxy due to these mergers. Major mergers as identified by the CAS parameters are defined as galaxy mergers where the mass ratio between the progenitors is $\mu = 1/4$ or less and so the amount of mass added to a galaxy during a merger must be similar to the original galaxy's mass. We take the ratio between progenitor masses to be an average of 1:1.5, as in \cite{conselice06b}, and we approximate the amount of stellar mass gained by a galaxy due to mergers by $\delta M_{\textup{merger}} \sim 1.65^{N_m} \times M_0$, where $N_m$ is the number of mergers and $M_0$ is the mass of the initial galaxy \citep{conselice06b}. Therefore, a galaxy of mass $M_{*} = 10^{9.5} M_{\odot}$ at $z = 3$ will gain a of mass $M_{*} = 1.3 \times 10^{10} M_{\odot}$ by $z = 0$, giving a relative mass increase from $z = 3$ to $z = 0$ of $\Delta M_{\textup{merger}}/M= 3.2$. A galaxy of initial mass $M_{*} = 10^{10.5} M_{\odot}$ will, on average, accumulate $M_{*} = 1.4 \times 10^{11} M_{\odot}$ by $z = 0$, giving a relative mass increase of $\Delta M_{\textup{merger}}/M= 4.1$.  The mass accretion rate, in terms of the amount of mass accreted from mergers per unit time, is  $\dot{M} \sim 10^{9} M_{\odot}$~Gyr$^{-1}$ at $M_{*} = 10^{9.5} M_{\odot}$, while for $M_{*} = 10^{10.5} M_{\odot}$ the mass accretion rate is a factor of ten times high, on average, giving $\dot{M} \sim 10^{10} M_{\odot}$~Gyr$^{-1}$. 

These are simply the average mass accretion rates per Gyr.  Naturally, this process is not smooth and mass will be added in discrete jumps during the 2.9 mergers at $z < 3$. However, this mass accretion rate due to mergers per galaxy is an average value for all galaxies that can be applied to a mass selected population.

\subsection{Comparison to Simulations}\label{subsec:compare_to_sim}

\begin{figure}[!ht]
\includegraphics[width=0.475\textwidth]{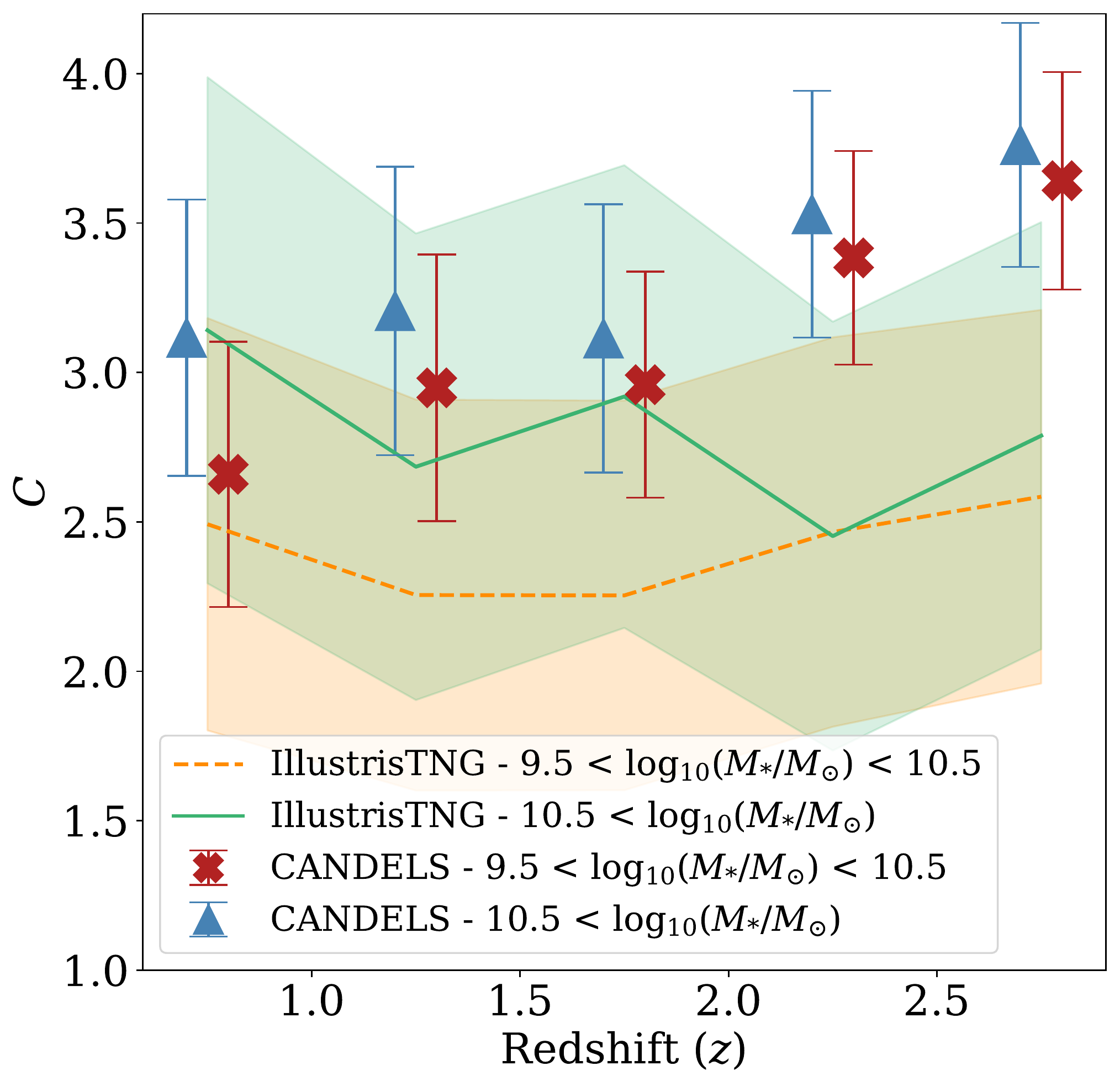}
\caption{Comparison of the evolution of the concentration for our results from CANDELS (corrected for redshift effects) and the galaxies within the IllustrisTNG simulations. As with Figure \ref{fig:conc_evol}, the CANDELS low mass ($10^{9.5}M_{\odot} < M_{*} <  10^{10.5}M_{\odot}$) galaxies are shown as blue triangles and the CANDELS high mass ($10^{10.5.5}M_{\odot} < M_{*}$) are shown as red crosses. The low mass IllustrisTNG simulations are shown as an dashed orange line and the high mass simulations are shown as a solid green line. The shaded regions indicate one standard deviation from the mean. The IllustrisTNG data is based upon high resolution images.}
\label{fig:sims_c}
\end{figure}

We compare our results to simulations within IllustrisTNG. Instead of using TNG300-1 as we do to determine the merger timescale, we use the higher resolution TNG50-1. This is done to ensure that morphologies are not affected by mass or spatial resolution, especially in the central regions. For TNG300-1, the radius containing $20\%$ of the light, $R_{20}$, is close to the softening length of the simulation for some galaxies, and as a result could impact the concentration measurement. From TNG50-1 we randomly select galaxies in the same mass range as our mass-selected sample, generate the mocks following the description in \S {\ref{subsec:tng}} and measure the CAS parameters using \textsc{Morfometryka}. In Figure \ref{fig:sims_c}, we show a comparison of the concentration evolution for our results from CANDELS and the IllustrisTNG simulations. The CANDELS results are corrected for redshift effects, as described in \S\ref{sec:corrections}. The IllustrisTNG simulations images are produced to have at least 3 times the resolution of the CANDELS images and have not been matched with \textit{HST} realism as the concentration should be universal for all resolutions. The resulting images are 640 $\times$ 640 pixels and the spectra have been shifted to coincide with the same filters as in Table \ref{tab:morph_filters}. As with Figure \ref{fig:conc_evol}, the CANDELS low mass ($10^{9.5}M_{\odot} < M_{*} <  10^{10.5}M_{\odot}$) galaxies are shown as blue triangles and the CANDELS high mass ($10^{10.5.5}M_{\odot} < M_{*}$) are shown as red crosses. The low mass IllustrisTNG simulations are shown as a dashed orange line and the high mass simulations are shown as a solid green line.  We find that the concentration of the simulations show very little evolution with redshift for both mass bins. This suggests that the galaxies are growing in size with the inner and outer regions growing at a similar rate whereas the trend in the CANDELS data suggests the galaxies are growing from the inside out. 

\begin{figure}[!ht]
\includegraphics[width=0.475\textwidth]{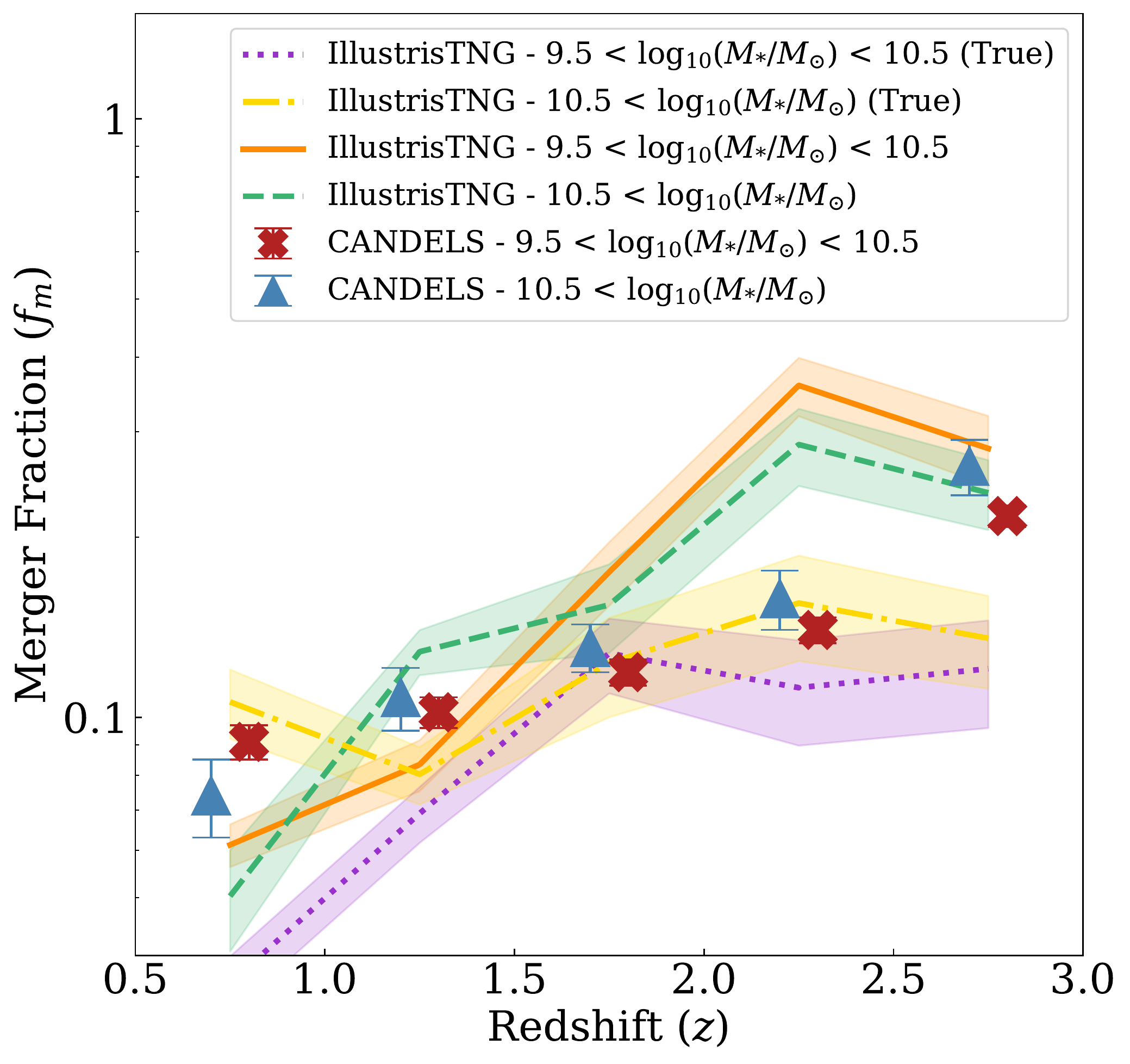}
\caption{Comparison of the evolution of the merger fraction for our results from CANDELS and the galaxies within the IllustrisTNG simulations. The IllustrisTNG data is based upon \textit{HST} matched mocks. As with Figure \ref{fig:conc_evol}, the CANDELS low mass ($10^{9.5}M_{\odot} < M_{*} <  10^{10.5}M_{\odot}$) galaxies are shown as blue triangles and the CANDELS high mass galaxies ($10^{10.5.5}M_{\odot} < M_{*}$) are shown as red crosses. The merger fraction evolution for the simulations determined using the asymmetry condition is shown as a solid orange line and a dashed green line for the low and high mass bins respectively. Also shown are the `true' merger fractions - the merger fraction determined based on the labels from the merger trees within the simulations and we select galaxies that are within 0.65 Gyr of a merging event - between the median and mean of the CAS time-scale distribution. These merger fractions are shown as a purple dotted line and a dot-dashed yellow line for the low and high mass bins respectively. Our results are consistent with both the asymmetry defined mergers and the `true' mergers from the simulations.}
\label{fig:sims_mf}
\end{figure}

We also compare the measured merger fractions within the simulations to the CANDELS data. For this, we randomly select $5\%$ of all available galaxies in TNG50-1 in this mass range. As this naturally selects more galaxies in the low mass bin than in the high mass bin, we further select more galaxies for the high mass bin until both bins have the same number of galaxies. The simulations in this case are calibrated on \textit{HST} resolution. This comparison is shown in Figure \ref{fig:sims_mf}. As for Figure \ref{fig:merge_evol}, the CANDELS data is shown as red crosses (low mass bin) and blue triangles (high mass bin). The IllustrisTNG merger fraction evolution determined using the same asymmetry condition as used for the CANDELS data for the low and high mass bins are shown as a solid orange line and a dashed green line respectively. We also determine the `true' merger fraction within the simulation by using the merger labels produced based on the merger trees. We ensure we only select major mergers by selecting mergers that have a mass ratio greater than 1/4. We also select only those galaxies that are within 0.65 Gyr of their merging event, a time between the median and mean of our CAS merger time-scale. These `true' merger fractions are shown as a purple dotted line and a dot-dashed yellow line for the low and high mass bins respectively. The shaded regions show the error on the merger fractions for the IllustrisTNG simulations. We find that both the `true' and asymmetry defined mergers are consistent with the observations. When selecting galaxies within a shorter timescale than the one chosen here, the number of galaxies in each redshift bin are too few and thus the `true' merger fractions decrease and do not agree very well with the asymmetry selected mergers from IllustrisTNG or the CANDELS data. This in part happens due to the time resolution between snapshots in IllustrisTNG, which is roughly $\Delta t \sim 0.16$ Gyr.  

We explore how many of galaxies within the Illustris simulation have $A > 0.35$ and also are considered to be a `true' major merger (with a mass ratio greater than 0.25). In this case we consider galaxies that are within 0.45 Gyr of a merger event, or 0.9 Gyr in total, to be `true' mergers in order to account for the time difference between snapshots within the simulation. We find that across all redshifts, an average of $\sim 37\%$ of the `true' major mergers lie above $A > 0.35$. This is slightly lower than the $\sim 50\%$ found by \cite{conselice06b}.

\section{Discussion}

In this paper we have presented a number of observed features of galaxy evolution, as viewed through the structural changes in galaxies over the past 11 Gyr. We carry this out through an analysis of all five CANDELS fields, examining the change in light concentration and asymmetry for these systems. As opposed to other studies of galaxy evolution at these epochs we take a largely structurally evolutionary view of galaxies, rather than one based on stellar populations. In this viewing of galaxies we are interested in the build up of stars in galaxies and how that affects the distribution of light within galaxies, rather than examining the type and distribution of stars themselves.

\subsection{Concentration}

We find that galaxies are more concentrated at higher redshifts when probing galaxy structure in the optical rest-frame.  This is an interesting result and is somewhat different from expectations, so it is important to examine the reasons for this. We test this result by examining the evolution of the concentration calculated using various measures of size. First, we correct the radii $R_{20}$ and $R_{80}$ (where $R_{20}$ is the radius containing 20\% of the total light within a galaxy and $R_{80}$ is the radius containing 80\% of the light) using the same method as described in \S\ref{sec:corrections} and use the mean of these corrected values to calculate the concentration using equation \ref{eq:C}. 

\begin{figure}[!ht]
\includegraphics[width=0.475\textwidth]{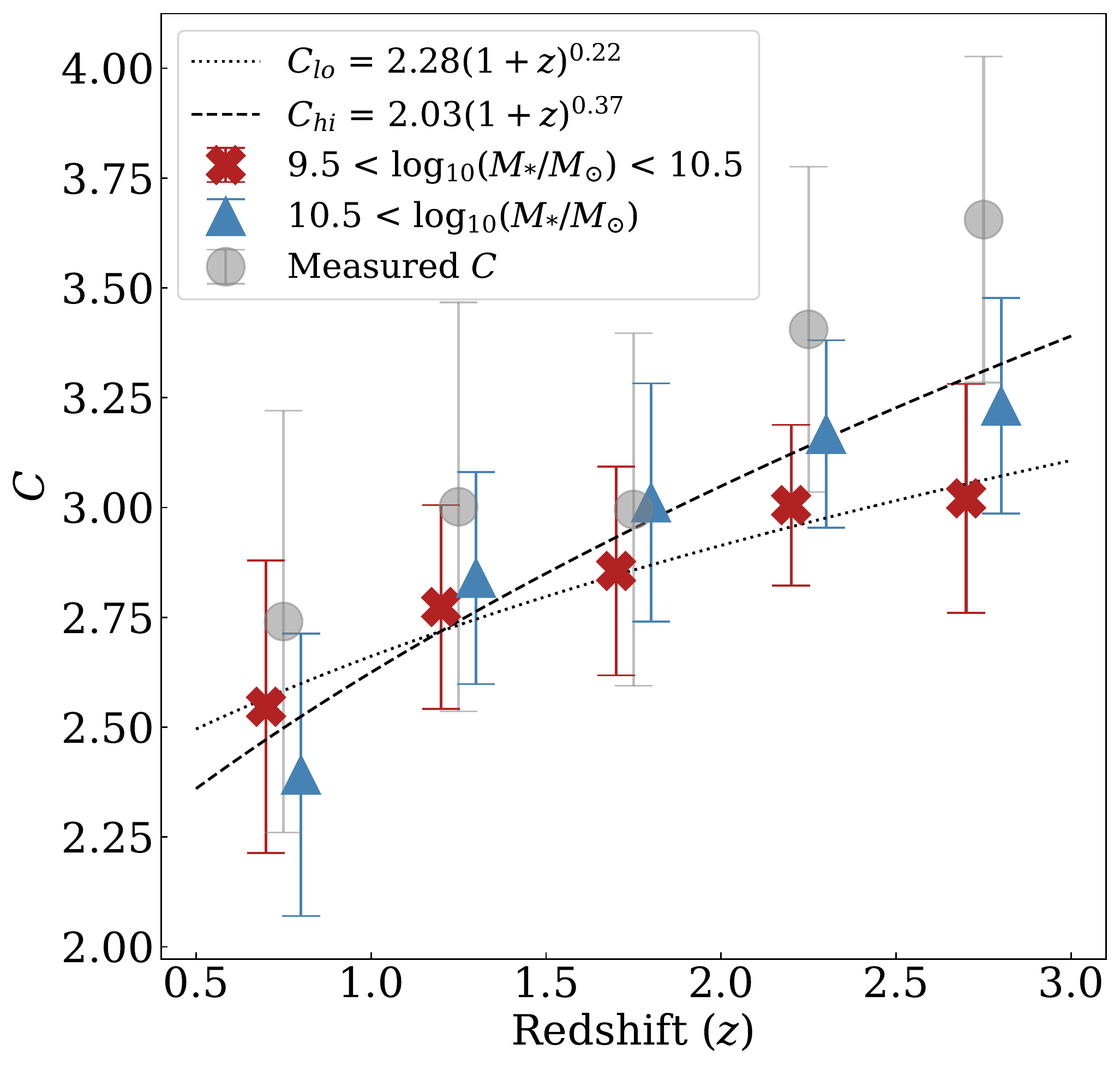}
\caption{Evolution of the concentration as calculated from the corrected $R_{20}$ and $R_{80}$ growth radii. The low mass bin is shown as red crosses and the high mass bin is shown as blue triangles. The grey circles represent the evolution of the measured concentration across all masses.}
\label{fig:CR20_R80}
\end{figure}

Examining the individual radii for our sample, we find that the evolution of the corrected $R_{80}$ values remain roughly constant with redshift whereas $R_{20}$, on average, increases with redshift. Therefore, we see a lower concentration at lower redshifts. We show the evolution of the concentration calculated using the corrected values of $R_{20}$ and $R_{80}$ in Figure \ref{fig:CR20_R80}. As in Figure \ref{fig:conc_evol}, the low mass bin is shown as red crosses and the high mass bin is shown as blue triangles. Also shown are power law fits to the concentrations, with the dotted line being for the low mass bin and the dashed line being for the high mass bin. For comparison, we show the evolution of the measured value of $C$ for all masses as grey circles. This decrease in concentration is consistent with the corrected concentrations shown in Figure \ref{fig:conc_evol}. 

We further explore this by examining the Petrosian radius, $R_{\textup{Petr}}(\eta)$. The Petrosian radius is defined to be the radius at which the surface brightness at a given radius is a specific fraction ($\eta$) of the surface brightness within that radius \citep[e.g.,][]{bershady00, conselice03, whitney19}. $\eta$ is given by:

\begin{equation}
\eta(r) = \frac{I(r)}{\left\langle I(r) \right\rangle}
\end{equation}

\noindent where $I(r)$ is the surface brightness at radius $r$ and $\left\langle I(r) \right\rangle$ is the mean surface brightness within that radius. Using this definition, $\eta = 0$ at large radii. We consider $R_{\textup{Petr}}(\eta=0.2)$ (the outer edge of a galaxy) and $R_{\textup{Petr}}(\eta=0.8)$ (the inner region of a galaxy). \cite{whitney19} show that $R_{\textup{Petr}}(\eta=0.2)$ grows more rapidly than $R_{\textup{Petr}}(\eta=0.8)$ for a mass-selected sample (where the masses lie in the range $10^{9}M_{\odot} < M_{*} <  10^{10.5}M_{\odot}$) across a redshift range of $1 < z < 7$ in the GOODS-North and GOODS-South CANDELS fields. So, one would expect the concentration at lower redshifts to be larger. In the same fashion as with the growth radii, we calculate the concentration of the galaxies in \cite{whitney19} with mass $\textup{log}_{10}(M_{*}/M_{\odot}) > 9.5$ using the Petrosian radii such that

\begin{equation}
    C_{\textup{P}} = 5 \times \textup{log}_{10}\left(\frac{R_{\textup{Petr}}(\eta=0.2)}{R_{\textup{Petr}}(\eta=0.8)}\right).
    \label{eq:C_petr}
\end{equation}

\begin{figure}[!ht]
\includegraphics[width=0.475\textwidth]{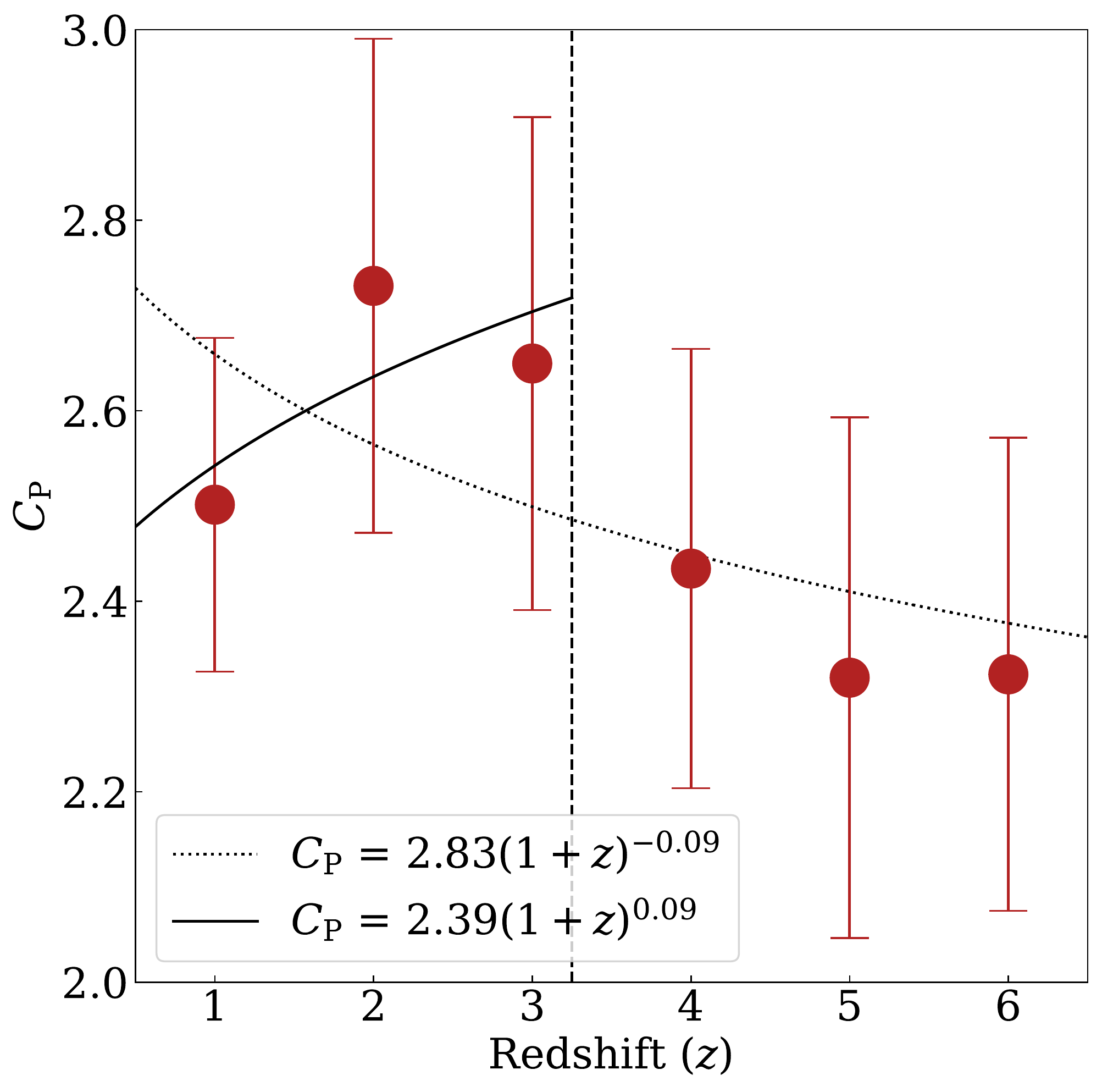}
\caption{Evolution of the concentration calculated using the Petrosian radii at $\eta = 0.2$ and $\eta = 0.8$ as per Equation \ref{eq:C_petr} for galaxies of mass $\textup{log}_{10}(M_{*}/M_{\odot}) > 9.5$ from \cite{whitney19}. We fit a power law to the whole redshift range, shown as a dotted line, and also to the redshift range we consider here (solid line). We find that $C_{\textup{P}}$ goes as $(1+z)^{-0.09 \pm 0.06}$ for the redshift range of $1 < z < 6$ and as $(1+z)^{0.09 \pm 0.09}$ for the smaller redshift range of $1 < z < 3$. The vertical dashed line indicates the redshift limit of this work. On average, the concentration decreases at higher redshift however when considering the lowest redshifts, the concentration increases with redshift.}
\label{fig:Cpetr}
\end{figure}

\noindent The resulting concentration evolution is shown in Figure \ref{fig:Cpetr}. We fit a power law to the entire redshift range from \cite{whitney19} (dotted line) and also the redshift range we consider in this paper (solid line). We find that $C_{\textup{P}}$ goes as $(1+z)^{-0.09 \pm 0.06}$ for the redshift range of $1 < z < 6$ and goes as $(1+z)^{0.09 \pm 0.09}$ for the smaller redshift range of $1 < z < 3$, which is within $2\sigma$ of being considered a flat relation. The vertical dashed line indicates the redshift limit of this work. Whilst on average over the redshift range of $1 < z < 6$, the concentration decreases with redshift, when we consider the lowest redshift bins, the concentration appears to increase with redshift, which is consistent with the results we find and show in Figure \ref{fig:conc_evol}. This is due to the fact that the inner radius appears to grow more rapidly compared to the outer radius within this small redshift range. We find the same result when considering galaxies with $\textup{log}_{10}(M_{*}/M_{\odot}) > 10.5$.  

Alternatively, it might be the case that as we go to lower redshifts, we are sampling more and more galaxies using a constant mass limit \citep{mundy15}. The result of this is that intrinsically lower halo mass galaxies will be entering our sample and these may have lower concentrations. We test this by considering how the concentration changes for a number density selection sample. Doing this at number densities of $3 \times 10^{-5}$ Mpc$^{-3}$ we find that the concentration decreases by a factor of $1.4 \pm 0.3$ from the highest redshift bin of $2.5 < z < 3$ to the lowest bin of $0.5 < z < 1$. This is comparable to the change seen in the mass-selected samples where the concentration decreases by a factor of $1.4 \pm 0.3$ for the lower mass bin and a factor of $1.2 \pm 0.2$ for the higher mass bin over the same redshift range. A comparison of the evolution for the mass-selected and number density-selected samples are shown in Appendix \ref{app:nd}. This suggests that the evolution we see in the concentration is real, and not due to newer galaxies being introduced to the lower redshift bins. 

Radius measurements of the inner regions can be inaccurate due to factors such as the size of the PSF. This could create inaccuracies in the concentration measurement. To ensure the concentration evolution we see is not an artefact of such issues, we also use $R_{50}$ and $R_{90}$ values in place of $R_{20}$ and $R_{80}$ in Equation \ref{eq:C}. We find that the evolution of $C_{59}$ exhibits similar behaviour to $C_{28}$, whereby the concentration increases by a factor of $\sim$1.5 over the redshift range $0.5 < z < 3$. We can therefore conclude that the evolution in concentration we see is real and not as a result of problems with the inner radius measurement.

We interpret a decreasing concentration with lower redshift as an indication that a small initial galaxy will eventually grow in size as a result of galaxy mergers, as well as galaxy accretion. This growth is occurring in such a way that the inner regions of a galaxy grow more rapidly in comparison to the outer regions at $z <3$.

\subsection{Merger Consistency}

Mergers can be identified in multiple ways such as pair statistics \citep[e.g.,][]{bundy09, man16, mundy17, duncan19}, CAS \citep[e.g.,][]{conselice09, bluck12}, Gini-$M_{20}$ \citep[e.g.,][]{lotz08, lotz10}, and Deep Learning \citep[e.g.,][]{ferreira20} with each of these methods yielding differing merger counts, and therefore fractions due to the different criteria. Using a power law fit, we find that the merger fraction goes as $(1+z)^{1.87 \pm 0.04}$ when fitting our data and setting the $z \sim 0$ merger fraction to a fixed value. This is shallower than that found by \cite{conselice09} who derive merger fractions using the CAS parameters for a sample of galaxies within the Extended Groth Strip and Cosmic Evolution Survey at $0.2 < z < 1.2$; they find a relation of $(1+z)^{2.3 \pm 0.4}$ using their data and by using a fixed value of the merger fraction at $z \sim 0$, the relation changes to $(1+z)^{3.8 \pm 0.2}$. \cite{ferreira20} use a convolution neural network trained on simulations from IllustrisTNG and applied to data from all five CANDELS fields and find an evolution of $(1+z)^{2.82 \pm 0.46}$, again slightly steeper than the results we find. \cite{duncan19} on the other hand find a shallower relation for their pair fractions of $(1+z)^{0.844^{+0.216}_{-0.235}}$ for galaxies with log$_{10}(M_{*}/M_{\odot}) > 10.3$. In general, the morphologically defined mergers give a higher merger fraction than those derived through pairs \citep{conselice09}.  However, this is not unexpected, as different methods of finding mergers will give different results for the fractions. 

Consistency is however expected when converting to merger rates and comparing these values.  Along with different methods giving different merger fractions, different selection criteria within the same sample also give varying results; \cite{man16} examine the pair fraction evolution up to $z \sim 3$ using the UltraVISTA/COSMOS $K_S$-band selected catalog along with 3DHST+CANDELS $H$-band imaging and explore the effect of using the flux ratio versus the mass ratio in the selection. When selecting via the stellar mass ratio, the merger fraction appears to increase with cosmic time in contradiction to the results we and others find. A selection via the $H$-band flux ratio however, the trend is similar to ours in that the merger fraction decreases. 

As with merger fractions, merging timescales vary quite drastically depending on the selection method. For example, \cite{lotz08} use GADGET N-Body hydrodynamical simulations and the radiative transfer code SUNRISE to create a sample of galaxy mergers. They use these to determine merging timescales for multiple methods of identifying mergers. Using the same asymmetry condition we use here, they determine a timescale of 0.94 $\pm$ 0.13 Gyr, whilst using a Gini and $M_{20}$ condition, they find the timescale to be 0.26 $\pm$ 0.10 Gyr. \cite{conselice08} who examine the merger rate up to $z \sim 3$ in the Hubble Ultra-Deep Field use a merger time-scale of 0.34 Gyr, determined using the asymmetry condition. These methods all assume a constant timescale with redshift, however using a close pair method of identifying mergers, \cite{snyder17} find that the timescale varies with redshift as $\tau \propto (1+z)^{-2}$. These timescales all differ from the constant timescale of $\tau = 0.56^{+0.23}_{-0.18}$ Gyr we find using the asymmetry criteria. 

Despite these differences in timescales and merger fractions for various methods of identifying mergers, merger rates are largely consistent with each other, as shown in Figure \ref{fig:merge_evol}. The merger rates determined using the asymmetry condition to identify mergers as we have done here are, within the errors, similar to merger rates determined using other methods such as probabilistic pair-counts \citep{duncan19} and finding mergers using Deep Learning \citep{ferreira20}. This is despite using different timescales for a merging event, showing that while different methods identify different numbers of mergers, our CAS method gives a consistent merger history for galaxies. The merger rates presented here have greater errors associated with them than previous work due to the consideration of the error in the merger timescale; the IllustrisTNG simulations used here to determine the merger timescale have an asymmetric distribution in the timescale and so we therefore use an asymmetric error. \cite{ferreira20} use a timescale based on their selection and as such, do not have an associated error. \cite{duncan19} calculate the errors on their merger fractions using the bootstrap technique. The timescale taken from \cite{snyder17} does not have an associated statistical error so when determining the fit (by convolving the merger fraction fits with the redshift dependent timescale), the merger rates only consider the error in merger fraction. Therefore, we have found that using three independent methods that we are able to measure a consistent merger rate at $z < 3$, and therefore have a firm idea of what the history of galaxy major mergers are within this redshift range.

\section{Conclusions}

Galaxy structure is one of the most fundamental ways in which galaxies and their formation and evolution can be understood, yet there have been few systematic studies of this property at higher redshifts. In this paper we collate the high resolution \textit{HST} imaging of  galaxies from all five fields of the CANDELS survey to determine the time evolution of galaxy structure to derive the processes for galaxy evolution. We combine these 16,778 galaxy structures, both visually based and quantitatively based, with stellar mass measurements and photometric redshifts to derive the evolution of galaxy structure. To aid our investigation of these morphologies we also utilise Illustris TNG300 simulations to calibrate timescales for morphological features and Illustris TNG50 for direct morphological comparisons with observations.

Our conclusions, based on a detailed non-parametric analysis of the structures of galaxies in all five CANDELS fields are as follows:

\begin{enumerate}
    \item We find that galaxies are distributed in a concentration-asymmetry plane up to $z \sim 3$ in a similar way as they do at lower redshifts. At these redshifts there is a continuum of parameters, suggesting unlike the colour-magnitude plane  galaxies on this relation evolve continuously over time, rather than change dramatically.  
    
    \item Based on dynamical simulations with the IllustrisTNG simulation, we show that the evolution of galaxies within the $C-A$ plane over this time period is largely driven by galaxy formation processes, including the accretion of gas as well as galaxy mergers.  
    
    \item We find that galaxies are both more asymmetric and more concentrated at higher redshifts. This demonstrates that galaxies are assembled in a way where material is coming in through galaxy formation processes that are relatively changing the inner portions of galaxy light quicker than the outer radii at $z < 3$.
    
    \item Using merger timescales from Illustris TNG300 we determine how the morphological timescales for mergers evolves with time. We find that for the CAS system of finding mergers, the mean merger timescale is $0.56^{+0.23}_{-0.18}$ Gyr. We also find, that unlike in the galaxy pair situation, this merger timescale does not evolve with time.
    
    \item Using these timescales we are able to determine the merger fraction evolution and the merger rate evolution up to $z \sim 3$. We find a good agreement between our merger rate calculations in comparison with previous work using pairs of galaxies and finding mergers based on Deep Learning. This is therefore a consistent merger history for galaxies as measured in three different ways.  We now have a firm and consistent measure of the galaxy major merger history at $z < 3$.
\end{enumerate}

Our results are the best measurements of the merger history using \textit{HST} imaging that will likely be possible unless a new large area survey in the near infrared is performed.  When Euclid and \textit{JWST} launch we will be able to expand these results in many ways.  We will be able to measure the merger history in a similar way up to $z \sim 6$, and at lower redshifts we will be able to probe down to lower masses to find the evolution of a fuller sampling of the galaxy population. Euclid will allow us to examine the merger history over 15,000 deg$^{2}$, allowing detailed merger histories to be measured.  The techniques and methods here can be used with both data sets, and others, to fully explore and measure this merger history, a principle part of the entire galaxy formation process.

\section{Acknowledgements}

This work is based on observations taken by the CANDELS Multi-Cycle Treasury Program with the NASA/ESA \textit{HST}, which is operated by the Association of Universities for Research in Astronomy, Inc., under NASA contract NAS5-26555. We thank the rest of the CANDELS team for their heroic work making their products and data available.  We acknowledge funding from the Science and Technology Facilities Council (STFC) to support this work.  LF acknowledges funding from the Coordena\c{s}\~{a}o de Aperfei\c{c}oamento de Pessoal de N\'{i}vel Superior - Brazil (CAPES).

\appendix 

\section{Number Density-Selected Sample} \label{app:nd}

In order to test whether the mass limits of our sample are creating an artificial trend, we instead select galaxies using a constant number density of $3 \times 10^{-5}$ Mpc$^{-3}$. This yields galaxies in the mass range $10^{10.7} M_{\odot} \leq M_{*} <  10^{11.5} M_{\odot}$. This method avoids problems with having a fixed stellar mass bin that galaxies can enter at lower redshift, compared with those that are at higher redshifts \citep[e.g.,][]{mundy15, ownsworth16}.  This has an effect such that by $z = 0$ only $\sim 5$ \% of galaxies with a high mass selection would have been within the same mass selection at $z > 2.5$.  However, by selecting through a constant co-moving number density we are able to remove this bias and always examine what are statistically the same number of massive galaxies at each redshift.   Because there are so many `new' galaxies in a mass-selected sample, it is possible that the relations we derive are biased by these new systems.

\begin{figure}[!ht]
\centering
\includegraphics[width=0.475\textwidth]{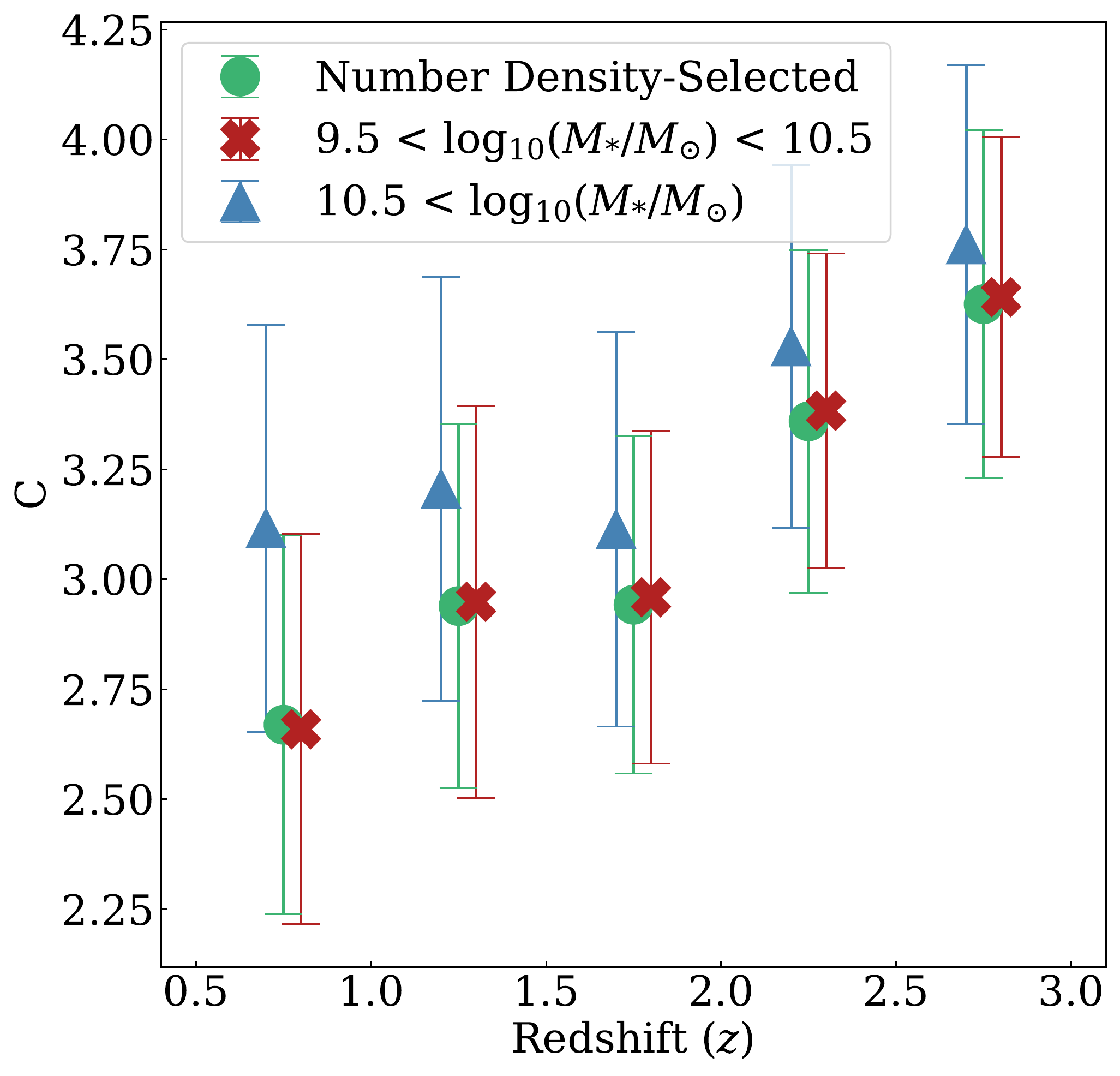}
\caption{A comparison of the evolution of concentration for a number density-selected sample (green circles) and a mass-selected sample. The number density is kept constant at $3 \times 10^{-5}$ Mpc$^{-3}$. As for Figure \ref{fig:conc_evol}, red crosses indicate the lower mass bin and blue triangles indicate the higher mass bin.  We find no significant difference in the evolution of the $C$ parameter using these different selection methods.}
\label{fig:nd_C}
\end{figure}

We determine the concentration evolution for this number density selected sample, and compare it to the evolution seen for the mass selected-sample. We find that the evolution for the number density-selected sample is very similar to that of the mass-selected sample, as shown in Figure \ref{fig:nd_C}. Green circles indicate the concentration for the number density-selected sample, red crosses indicate the lower mass bin, and blue triangles the higher mass bin.   Therefore our finding of a reduction in concentration using the mass selected sample is not due to new galaxies entering the criteria.

\begin{figure}[!ht]
\centering
\includegraphics[width=0.475\textwidth]{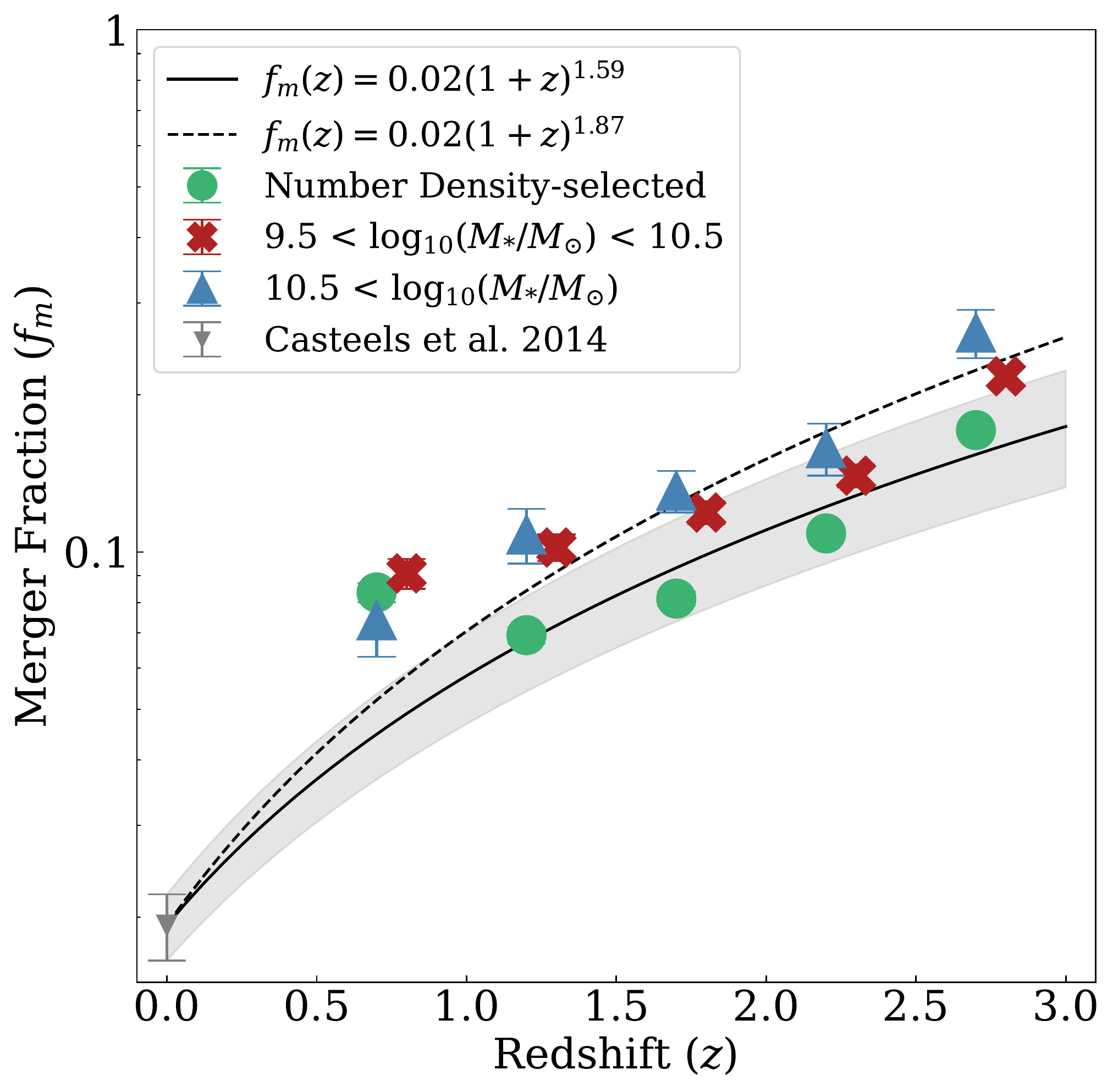}
\caption{A comparison of the evolution of the merger rate for a number density-selected ($n = 3 \times 10^{-5}$ Mpc$^{-3}$) sample of galaxies (green circles), low mass galaxies (red crosses), and higher mass galaxies (blue triangles). The fit to the number density-selected sample is shown by the solid line and the fit to the mass-selected sample is shown by the dashed line. The known local value of the merger fraction from \cite{casteels14} is from a mass-selected sample.}
\label{fig:nd_mf}
\end{figure}

As a test for another key result, we examine how the merger fraction evolution would occur by using a selection based on this number density selection of $3 \times 10^{-5}$ Mpc$^{-3}$.  This is shown in Figure \ref{fig:nd_mf} where we find that the merger fraction is slightly lower for all but the final lowest redshift bin.  However, the fitted merger fraction evolution and thus the corresponding merger rate is statistically very similar and does not give significantly different merger histories for galaxies.  Thus, our result using a constant mass selection is a viable way in which to trace the evolution of the massive galaxy population, and our results are robust to galaxy selection methods.

\bibliography{bibliography}{}
\bibliographystyle{aasjournal}



\end{document}